\documentclass[twocolumn]{aastex631}
\usepackage{amsmath}
\usepackage{mathrsfs} 

\newcommand{\MJ}[1]{#1}

\newcommand{\hGpc}{{\ifmmode{\,h^{-1}{\rm Gpc}}\else{$h^{-1}$Gpc}\fi}}

\newcommand{\hMpc}{{\ifmmode{\,h^{-1}{\rm Mpc}}\else{$h^{-1}$Mpc}\fi}}
\newcommand{\hkpc}{{\ifmmode{\,h^{-1}{\rm kpc}}\else{$h^{-1}$kpc}\fi}}
\newcommand{\hMsun}{{\ifmmode{\,h^{-1}{\rm {M_{\odot}}}}\else{$h^{-1}{\rm{M_{\odot}}}$}\fi}}
\newcommand{\Msun}{\,\rm {M_{\odot}}}
\newcommand{\Mstar}{{\ifmmode{\,M_{*}}\else{$M_{*}$}\fi}}
\newcommand{\Mhalo}{{\ifmmode{\,M_{\rm halo}}\else{$M_{\rm halo}$}\fi}}
\newcommand{\ltsima}{$\; \buildrel < \over \sim \;$}
\newcommand{\gtsima}{$\; \buildrel > \over \sim \;$}
\newcommand{\lsim}{\lower.5ex\hbox{\ltsima}}
\newcommand{\gsim}{\lower.5ex\hbox{\gtsima}}

\newcommand{\theth}{\textsc{The Three Hundred}}
\newcommand{\ahf}{\textsc{AHF}}

\newcommand{\gadgetx}{\textsc{GADGET-X}}
\newcommand{\simba}{\textsc{GIZMO-SIMBA}}
\newcommand{\gadgetmusic}{\textsc{Gadget-MUSIC}}

\begin{document}

\title{Intrinsic mass-richness relation of clusters from THE THREE HUNDRED hydrodynamic simulations}

\author{Mingjing Chen}
\email{mingjing@mail.ustc.edu.cn}
\affiliation{CAS Key Laboratory for Research in Galaxies and Cosmology, Department of Astronomy, University of Science and Technology of China, Hefei, Anhui, 230026, People’s Republic of China}
\affiliation{School of Astronomy and Space Science, University of Science and Technology of China, Hefei, Anhui, 230026, People’s Republic of China}

\author{Weiguang Cui}
\thanks{Talento-CM fellow}
\email{weiguang.cui@uam.es}
\affiliation{Departamento de F\'{i}sica Te\'{o}rica, Universidad Aut\'{o}noma de Madrid, M\'{o}dulo 15, E-28049 Madrid, Spain}
\affiliation{Centro de Investigaci\'{o}n Avanzada en F\'isica Fundamental (CIAFF), Facultad de Ciencias, Universidad Aut\'{o}noma de Madrid, 28049 Madrid, Spain}
\affiliation{Institute for Astronomy, University of Edinburgh, Royal Observatory, Edinburgh EH9 3HJ, UK}

\author{Wenjuan Fang}
\email{wjfang@ustc.edu.cn}
\affiliation{CAS Key Laboratory for Research in Galaxies and Cosmology, Department of Astronomy, University of Science and Technology of China, Hefei, Anhui, 230026, People’s Republic of China}
\affiliation{School of Astronomy and Space Science, University of Science and Technology of China, Hefei, Anhui, 230026, People’s Republic of China}

\author{Zhonglue Wen}
\email{zhonglue@nao.cas.cn}
\affiliation{CAS Key Laboratory of FAST, NAOC, Chinese Academy of Sciences, Beijing 100101, People’s Republic of China}
\affiliation{National Astronomical Observatories, Chinese Academy of Sciences, 20A Datun Road, Chaoyang District, Beijing 100101, People’s Republic of China}

\submitjournal{\apj}
\date{\today}
\received{February 07, 2024}
\accepted{March 28, 2024}

\begin{abstract}
The main systematics in cluster cosmology is the uncertainty in the mass-observable relation.
In this paper, we focus on the most direct cluster observable in optical surveys, i.e. richness, and constrain the intrinsic mass-richness (MR) relation of clusters in THE THREE HUNDRED hydrodynamic simulations with two runs: GIZMO-SIMBA and GADGET-X. 
We find that modeling the richness at fixed halo mass with a skewed Gaussian distribution yields a simpler and smaller scatter compared to the commonly used log-normal distribution.
Additionally, we observe that baryon models have a significant impact on the scatter, while exhibiting no influence on the mass dependence and a slight effect on the amplitude in the MR relation.
We select member galaxies based on both stellar mass $M_\star$ and absolute magnitude $\mathscr{M}$. We demonstrate that the MR relation obtained from these two selections can be converted to each other by using the $M_\star-\mathscr{M}$ relation.
Finally, we provide a 7-parameter fitting result comprehensively capturing the dependence of the MR relation on both stellar mass cutoff and redshift.

\end{abstract}


\section{introduction}\label{sec01}
Galaxy clusters (hereafter clusters for simplicity), as the largest gravitationally bound structures in the Universe, hold significant importance in both cosmology and astrophysics \citep[see][etc. for reviews]{Kravtsov2012,Allen2011,Wechsler2018}.
Accurate measurement of cluster mass is one of the most crucial steps for these studies \citep{Pratt2019}.

Different methods can be used to determine individual cluster's mass.
The simplest and oldest method is dynamical analysis, using galaxy velocity dispersion with the assumption of dynamical equilibrium \citep{Zwicky1937,Li2021}.
X-ray observations estimate cluster mass through gas density and temperature profiles with the hydrostatic equilibrium assumption \citep[see][for example]{Ansarifard2020,Pearce2020,Gianfagna2021}.
 On top of that, the strong and weak lensing signals from shape distortions of background galaxies provide a most direct and powerful method to measure the cluster mass \citep[e.g.][]{Meneghetti2010,Okabe2016}.
In general, these different methods yield consistent results in previous studies \citep[e.g.][]{Lewis1999}.

However, these methods require high-quality or long-term spectral observations, restricting accurate measurements to only a small number of clusters.
To overcome this limitation and to obtain a large number of cluster masses extending to high redshift which is important for cosmology, the cluster mass-observable relation is commonly employed, i.e. estimating the masses of a cluster sample using more easily accessible observables as mass proxies.
This approach has been widely utilized in cosmological research after being calibrated with direct measures of cluster masses, such as weak lensing \citep[e.g.][]{McClintock2019}, or through self-calibration when constraining cosmological parameters \citep[e.g.][]{Oguri2011}.

Different mass proxies are utilized in different surveys.
In X-ray surveys, commonly used mass proxies include the gas mass, gas temperature, gas luminosity in different X-ray bands or integrated \citep[e.g.][]{Mulroy2019,Babyk2023}. In Sunyaev-Zel'dovich (SZ) surveys, the projected integrated SZ flux is usually used \citep[e.g.][]{Planck2016SZ}. Optical surveys make use of observables such as richness, optical luminosity and galaxy overdensity \citep[e.g.][]{Pearson2015} as mass proxies. 
Compared to X-ray and SZ surveys, optical surveys have a larger field of view and can easily extend to higher redshift with bigger signal-to-noise ratios. Multi-wavelength bands in optical surveys are generally available which can provide photometric redshift if the spectroscopic redshift is not available. Albeit a slightly large error, this enables the detection of clusters to higher redshifts. Consequently, a large sample of clusters spanning a wide range of mass and redshift can be constructed \citep[e.g.][]{whl12,Rykoff2014,wh21}.
Among these optical observables, richness is the most direct one and exhibits a small scatter \citep{Old2014,Old2015,Pearson2015}, which is of utmost importance for cosmological constraints.
Although cluster member identification suffers from foreground and background contamination, as well as these interlopers \citep{Wojtak2018}, which introduce uncertainties in richness. Advancements in cluster finding techniques have enabled richness to remain a reliable mass proxy with low scatter \citep{Rykoff2012,Rykoff2014}.

Numerous articles have been devoted to constraining the mass-richness relation, hereafter MR relation. For instance, some studies are based on X-ray measurements, such as \cite{Capasso2019} using the ROSAT All-Sky Survey and \cite{Chiu2023} using the extended ROentgen Survey with an Imaging Telescope Array (eROSITA), and some studies based on SZ measurements, like \cite{Saro2015} and \cite{Bleem2020}, utilizing the South Pole Telescope (SPT). 
Additionally, studies from optical surveys, such as \cite{Murata2018} and \cite{Simet2017} using the Sloan Digital Sky Survey (SDSS) redMaPPer clusters, \cite{Murata2019} utilizing the Subaru Hyper Suprime-Cam (HSC), and \cite{Costanzi2021} employing the Dark Energy Survey (DES), are based on the weak lensing measurements of clusters.

These studies typically employ a power-law model to describe the MR relation. Most of them report consistent dependencies on mass, aligning with the predictions of self-similarity \citep{Kaiser1986}. However, discrepancies arise when it comes to the redshift dependence. \cite{Andreon2014} and \cite{Saro2015} argue that the data is consistent with no redshift evolution within $1\sigma$, while \cite{Capasso2019} demonstrates a strong negative evolution trend.
Regarding the treatment of the richness probability distribution, most studies adopt a log-normal distribution, albeit employing different formulas for the scatter. Some studies \citep{Murata2018,Murata2019} take it as a linear function of  the logarithm of mass and redshift to account for observational effects. Others \citep{Capasso2019,Bleem2020,Costanzi2021} model it as a Poisson term plus an intrinsic scatter term, separately accounting for projection effects. 

Few articles investigate thoroughly the intrinsic MR relation from a theoretical standpoint. In this work, we aim at such a study. Specifically, we employ a power-law model for the MR relation, similar to previous studies, but delve deeper to examine its dependencies on redshift, limit of galaxy stellar mass or magnitude for member galaxy selection.
The most important aspect of our work lies in the choice for the richness probability distribution. Instead of employing a simple log-normal distribution as in previous studies, we utilize a skewed Gaussian distribution with a scatter based on the Halo Occupation Distribution (HOD) model \citep{Jiang2017}. Notably, this choice results in a mass-independent intrinsic scatter.
Our work is based on two different hydro-simulations starting from the same initial conditions but different baryon models \citep{Cui2018,Cui2022},. 
The outcomes of this study can improve our understanding of the MR relation, and contribute to accurate modeling approaches, which, in turn, can hopefully reduce the scatter in the MR relation and  ultimately tighten the constraints on cosmological parameters.

This paper is organized as follows. In \autoref{sec02}, we introduce \theth\ on which our analysis is based. \autoref{sec03} describes our model for the MR relation with a skewed Gaussian distribution for the richness. In \autoref{sec04}, we present the main results for both selection of galaxies based on galaxy stellar mass and on magnitude. \autoref{sec05} involves comparing our results with other prescriptions for the richness distribution, as well as including the dependences on the stellar mass limit and redshift. We also make comparison with other findings from the literature. Finally, we summarize and conclude in \autoref{sec06}.

\section{the \MJ{simulated} data}\label{sec02}

    \subsection{\theth}\label{sec21}

    \theth (hereafter \MJ{THE300}) \citep{Cui2018} performs hydrodynamic cosmological zoom-in re-simulations in 324 selected cluster regions.
    These regions are spherical with a radius of 15 \hMpc, centered around the 324 most massive clusters extracted from the MultiDark Planck 2 simulation (MDPL2) \citep{Klypin2016}.
    MDPL2 is a dark matter-only N-body simulation with a comoving length of 1 \hGpc, using $3840^3$ dark matter particles of mass $m_\text{DM}=1.5\times 10^9\hMsun$, and adopts cosmological parameters from \cite{Planck2016}.
    
    The re-simulation process initializes the parent dark matter particles into dark matter $m_\text{DM}=1.27\times 10^9\hMsun$ and gas components  $m_\text{gas}=2.36\times 10^8\hMsun$, then conduct three different baryonic codes: GADGET-MUSIC \citep{Sembolini2013}, \gadgetx\ \citep{Rasia2015}, and \simba\ \citep{Dave2019,Cui2022}. Thanks to \MJ{THE300}'s unique setups, for example, the large surrounding area of the central cluster, the filamentary structures connecting to the cluster are studied \citep{Kuchner2020, Rost2021, Kuchner2021,Rost2024}; the large sample of clusters permits statistical studies on cluster profiles \citep{Mostoghiu2019, Li2020, Baxter2021}, \MJ{back-splash} galaxies \citep{Arthur2019, Haggar2020, Knebe2020}, cluster dynamical state \citep{DeLuca2021, Capalbo2021, Zhang2022, Li2022}, lensing studies \citep{Vega-Ferrero2021, Herbonnet2022, Giocoli2023} and cluster mass \citep{Li2021, Gianfagna2023}; it is further used for the machine learning studies \citep{deAndres2022, deAndres2023, Ferragamo2023}.
    
    In this paper, we only focus on the results from \gadgetx\ and \simba\ runs. \MJ{We do not consider \gadgetmusic\ due to its lack of AGN feedback, which results in an overabundance of massive galaxies compared to actual observations \citep[see Fig.7 in][]{Cui2018}. That is unrealistic and will significantly alter the MR relation with a higher galaxy stellar mass cut.} For details of the two simulation models \MJ{we study}, we refer to \cite{Cui2018,Cui2022} for the general comparisons and the references therein for more information on the detailed implementation of the baryon models. Here, we briefly mention that the former is mostly calibrated based on gas properties, which present better agreement to the observation in gas properties, such as density/temperature profiles \citep{Li2020, Li2023}. While the latter is calibrated based on the stellar properties as described in \cite{Cui2022}. Nevertheless, the cluster's global properties are very similar.
    
    \begin{figure*}
	\centering
	\includegraphics[width=\linewidth]{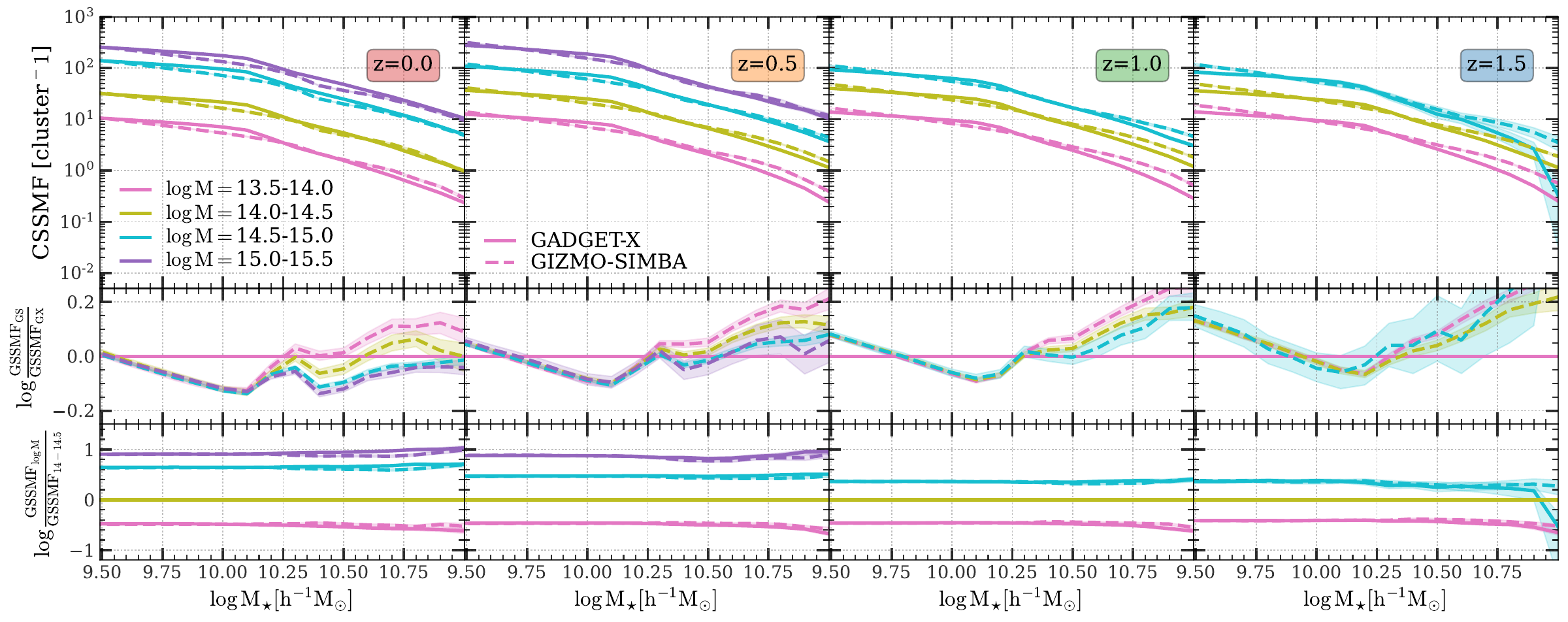}
	\caption{Cumulative satellite galaxy stellar mass function (CSSMF) per cluster, from the \gadgetx\ (solid lines) and the \simba\ (dashed lines) simulations  at different redshifts and for different halo mass bins.
    The shaded regions show 68 percent confidence intervals from bootstrap resampling.
    From left to right, each column corresponds to redshift $z=[0,0.5,1.0,1.5]$, respectively, and different colors represent different halo mass bins as indicated in the legend.
    The second row shows the difference in logarithmic CSSMF between the two simulations.
    The third row represents the difference in logarithmic CSSMF between a given halo mass bin and the one displayed as the yellow line.
    }
	\label{fig:cumuSSMF}
    \end{figure*}
    \subsection{the halo and galaxy catalogues}\label{sec22}		
    
    We utilize four snapshots, corresponding to redshifts $z=[0,0.5,1,1.5]$, for all the halos within the 324 cluster regions.
    
    Within each region, halos are first identified by \ahf \citep{Knollmann_Knebe2009}, a halo finder based on the spherical overdensity (SO) algorithm. We only consider halos with mass $M=M_{200c}>1\times 10^{13}\hMsun$, where $M_{200c}$ is defined as the mass enclosed within a radius $R_{200c}$ where the average density is 200 times the critical density at the redshift of the halo.

    Galaxies within these halos are further identified by Caesar, based on a 6-dimensional friends-of-friends (6DFOF) algorithm. Considering the resolution, we only include galaxies with stellar mass $M_\star \geq 10^{9.5} \hMsun$, to ensure at least 10 stellar particles per galaxy. Additionally, \MJ{we} also exclude those host halos which are contaminated by low-resolution particles. 

    In \autoref{fig:cumuSSMF}, we show the cumulative satellite stellar mass functions (CSSMF), which represent the total number of satellite galaxies with stellar masses greater than $M_\star$ per cluster. The CSSMFs are derived from all the selected halos binned in different halo masses (different color lines). Different redshift results are shown in different columns. 
    It is interesting to see that the CSSMFs scale almost perfectly with host halo mass as shown in the bottom row at all galaxy stellar masses, albeit only little variations at the massive galaxy stellar mass end. Though the lines are still parallel to the horizontal golden line, the exact constant values seem to vary (get closer to the golden line) slightly from low to high redshift, $z=1.5$. The two simulations are also in very perfect agreement, except for the tiny change at $M_\star \gtrsim 10^{10.25}\hMsun$. This suggests that the slope of the MR relation will be quite similar between the two simulations but decreases weakly with the redshift. 
    
    The absolute differences between \gadgetx\ and \simba\ are shown in the middle row, which clearly depend on the galaxy's stellar mass. And this dependence is also tangled with the host halo masses at higher galaxy stellar mass, $M_\star \gtrsim 10^{10} \hMsun$. This dependence further evolves with redshift as well: Although the first deep's position -- at $\sim 10^{10.1} \hMsun$ corresponding to the crossing point in Fig. 8 in \cite{Cui2022} -- is more or less stable at different redshifts, the relative difference curves shift up as redshift increasing to $z=1.5$; The middle peak at around $\sim 10^{10.3} \hMsun$ at $z=0$ is getting weaker and almost disappeared at higher redshift. This mostly connects to the relative difference between \gadgetx\ and \simba\ on the normalization parameter of the MR relation, while this normalization parameter is determined by the values of the CSSMFs which are presented on the top row of \autoref{fig:cumuSSMF}. 
    
    It is interesting to note that there is a small increase of CSSMF within the same halo mass bin tracking back to higher redshifts. This could be caused by several reasons, e.g. the pseudo halo evolution resulting from the fact that we are using $R_{200c}$; the halo evolution which changes its density profile either because of accretion or merger. We made a simple comparison between the simulated and analytical $R_\zeta \equiv R_{200c} (z=1)/R_{200c}(z=0)$ with a concentration parameter from \cite{Duffy2008} and found that the simulated $R_\zeta$ is larger than the analytical one, which suggests that the halo evolution plays a major role in this CSSMF in agreement with \cite{Ahad2021}. This can be simply explained as the halos are still in the formation process through mergers at high redshift, which can also be viewed as the relaxation fraction of the cluster's dynamical state drops along the redshift \citep[see][for example]{DeLuca2021}.
    
    Magnitudes of the galaxies are also provided by Caesar, using the flexible stellar population synthesis code FSPS \citep{Conroy2009, Conroy2010}. Dust obscuration is also taken into account in this study for \simba, because it has the dust model included \citep[see][]{Li2019}. However, there is no dust attenuation for \gadgetx. We don't include that for \gadgetx\ for two reasons: (1) there is very little dust in these cluster satellite galaxies, which has especially been verified in \simba; (2) simple dust attenuation laws, such as \cite{Charlot_Fall2000}, will only affect the magnitude systematically for all the galaxies at a particular band. So, it will have minimal effect on our results \MJ{. For example, at $z=0$, only 4.6\% of galaxies exhibit a fractional difference greater than 0 between the CSST i band absolute magnitudes considering dust and without considering dust, while, only 2.06\% of galaxies have a fractional difference greater than 0.01}. More complex models require a lot of assumptions, which may not be \MJ{worth it given that dust contributes little in the cluster environment suggested by} \simba.
    Our analysis focuses mainly on the ongoing and upcoming large optical surveys, namely the Chinese Space Station Telescope \citep[CSST,][]{Zhan2011}, and Euclid \citep{Laureijs2011}. Specifically, we consider the CSST i-band and z-band magnitudes, as well as the Euclid h-band magnitude in this study. We note here that the simulation used in this paper may not be able to reach the Euclid limits at low redshift \citep[see][]{Jimenez2023}. However, this is not a major concern for our MR relation study, because (1) we are studying different magnitude/stellar mass limits, above which all galaxies are included; (2) our results have a better convergence with low limits, such that it would be safe to extend our conclusions/fitting parameters to an even lower limit.

\section{method}\label{sec03}
\subsection{model}\label{sec31}

In the absence of non-gravitational physical processes during cluster formation, cluster scaling relation will follow a self-similar model prediction \citep{Kaiser1986}. The self-similar model predicts power-law scaling relations, which have been used in many simulations and observational studies.
\begin{equation}\label{eq:lnRichness}
	\left<\ln \lambda|\ln M\right>=A+B \ln \left(\frac{M}{M_{ {piv }}}\right),
\end{equation}
where $\lambda$ is the optical richness defined in the last section, $A$ is the normalization, $B$ is the slope with respect to the halo mass $M$, and $M_{piv}=3\times 10^{14}\hMsun$ is a pivot mass scale. 

We adopt forward modeling for the probability distribution function of optical richness for halos with a given mass $P( \lambda |  M)$.
The corresponding backwards $P( M | \lambda)$ has also been studied in many works \citep[e.g.][]{Simet2017}. The former allows for a more direct comparison of the model prediction with the measurements, while the latter is more suitable for inferring halo mass from observables. These two can be converted into each other by using the halo mass function \citep{Evrard2014}. Note that, modeling the $P( M | \lambda)$ is different from modeling the $P( \lambda |  M)$. This is because the $M$ in observation is subject to many systematics. Directly transferring from $P( \lambda |  M)$ to  $P( M | \lambda)$ needs Bayes theorem:
$$  P(M|\lambda) = \frac{P(\lambda|M)P(M)}{P(\lambda)}, $$ where $P(M)$ is related to the halo mass function. \cite{Evrard2014} gave an approximate solution: if $P( \ln \lambda | \ln M)$ is Gaussian with a scatter $\sigma_{\ln\lambda}$, $P( \ln M |\ln \lambda)$ will be Gaussian with a scatter $\sigma_{\ln M}=\sigma_{\ln\lambda}/B$ in the first order assuming $P(M)$ is simple power law and $P(\lambda)$ is a constant.

Typically, $P( \lambda |  M)$ is modeled as a log-normal distribution  \citep{Murata2018,Murata2019}. However, this form exhibits a negative skewness \citep{Anbajagane2020}, which is also expected from the HOD model.
In the HOD model, galaxies are categorized as central and satellite galaxies $\lambda=\lambda^\text{cen}+\lambda^\text{sat}$. The latter follows a sub-Poisson distribution at small occupation numbers and a super-Poisson distribution at large numbers \citep{Jiang2017}.
In the mass range we selected later, there is always a central galaxy with $\lambda^\text{cen}=1$, and the distribution for satellite galaxies is chosen to be super-Poisson because we are interested in galaxy clusters.

We model the deviation from Poisson as a Gaussian distribution with scatter $\sigma_\text{I}$ \citep{Costanzi2019}, which represents halo-to-halo variations influenced by the large-scale environments \citep{Mao2015}. Specifically, the richness can be written as $\lambda=\lambda^\text{cen}+\lambda^\text{sat}=1+\Delta^\text{Poisson}+\Delta^\text{Gauss}$, where  $\Delta^\text{Poisson}$ follows a Poisson distribution with a mean value of $\left< \lambda^{\text {sat }} \right>$, and $\Delta^\text{Gauss}$ follows a Gaussian distribution with a mean of $0$ and a scatter of $\sigma_\text{I}$.

\begin{figure}
	\centering
	\includegraphics[width=\linewidth]{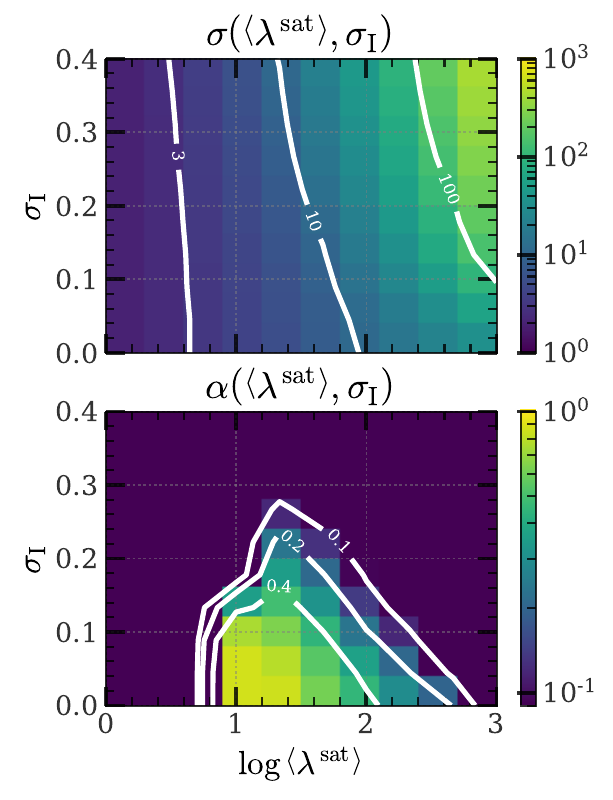}
	\caption{Two-dimensional interpolation tables for $\sigma $(upper panel) and $\alpha $(lower panel) as a function of $\{\left< \lambda^{\text {sat }} \right>,\sigma_\text{I} \}$.}
	\label{fig:interp2d}
\end{figure}

To obtain the probability distribution $P(\lambda)$, we sample $\lambda$ $10^6$ times for each $\{\left< \lambda^{\text {sat }} \right>,\sigma_\text{I} \}$. Then, we fit $P(\lambda)$ with a skewed Gaussian distribution by calibrating the parameters $\{  \sigma, \alpha  \}$:
\begin{equation}\label{eq:Probability}
	P\left(\lambda \mid M\right)=\frac{1}{\sqrt{2 \pi \sigma^{2}}}
	\\
	 \mathrm{e}^{-\frac{\left(\lambda-\left\langle\lambda^{\text {sat }}\right| M\right)^{2}}{2 \sigma^{2}}} 
	 \\
	 \operatorname{erfc} \left[-\alpha \frac{\lambda -\left\langle\lambda^{\text {sat }} \mid M\right\rangle}{\sqrt{2 \sigma^{2}}}\right],
\end{equation}
where $\left< \lambda^{\text {sat }} \right>=\exp \left<\ln \lambda\right>-1$.
For the subsequent calculations, we employ two-dimensional interpolation tables that relate $\{\left< \lambda^{\text {sat }} \right>,\sigma_\text{I} \}$ to the corresponding values of $\{  \sigma,\alpha  \}$ as shown in \autoref{fig:interp2d}.

\begin{figure}
	\centering
	\includegraphics[width=\linewidth]{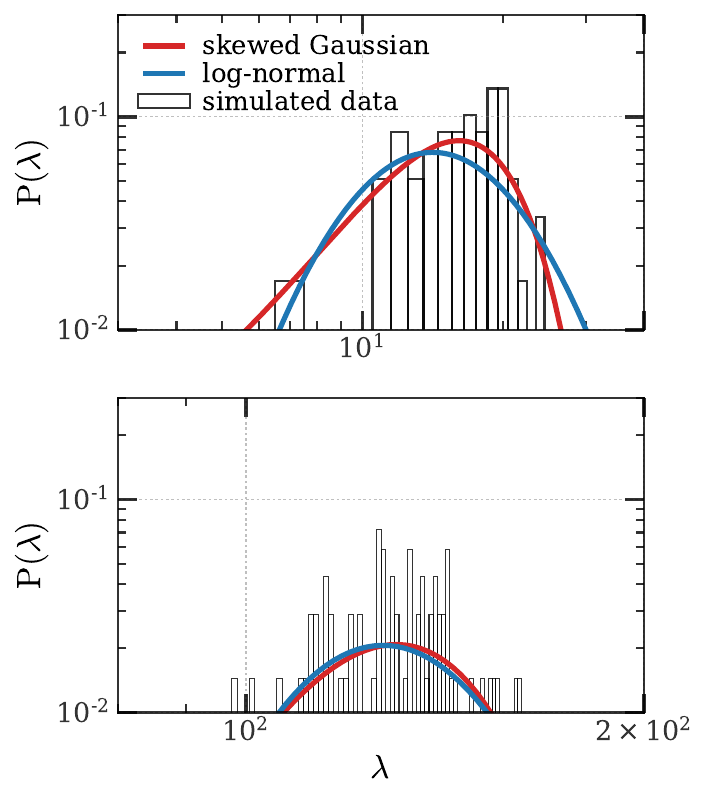}
	\caption{Richness distribution (black) for \gadgetx\ at $z=0$, as well as two fitting probability functions: skewed Gaussian (red) and log-normal (blue) function.
		The upper and lower panels correspond to mass bin $\log M[\hMsun]=[13.9,13.95]$ and $\log M[\hMsun]=[14.8,14.85]$, respectively.}
	\label{fig:DataDistribution}
\end{figure}

\autoref{fig:DataDistribution} shows an example utilizing this skewed Gaussian function to fit the richness probability distribution from the \gadgetx\ data in two mass bins at $z=0$, while also employing the commonly used log-normal function for comparison. The richness here is defined as the count of all member galaxies in the catalogue described in \autoref{sec22}.
The former demonstrates better incorporation of low richness values, while both exhibit greater consistency in the larger mass bin $\log M[\hMsun]=[14.8,14.85]$. Additionally, regardless of the mass bin, the residual of the former is consistently lower than that of the latter: $2.48<3.64$ for $\log M[\hMsun]=[13.9,13.95]$ and $15.53<15.96$ for $\log M[\hMsun]=[14.8,14.85]$.
More comparisons for different galaxy selections and for \simba\ are shown in the \autoref{sec:app1}.

 However, these two panels are fitted separately, which means that the mass dependence of the scatter is not taken into account. 
 For the scatter of the skewed Gaussian distribution $\sigma_\text{I}$, we will subsequently demonstrate that it exhibits no mass dependence. While for the scatter of the log-normal distribution, there is a widely used form \citep{Capasso2019,Bleem2020,Costanzi2021}:
\begin{equation}\label{eq:sigmaIG}
	\sigma_{\ln \lambda}^2=\sigma_\text{IG}^2+ \left(e^{\left< \ln \lambda \right>} -1 \right)/e^{2\left< \ln \lambda \right>}, 
\end{equation}
i.e., the sum of a constant intrinsic scatter with a Poisson-like term.
This form incorporates the mass dependence through the Poisson term, which is also motivated by the super-Poisson distribution in the HOD model. However, compared to our approach, it simplifies this assumption, resulting in an extra mass dependence. We will demonstrate this from two perspectives.

\begin{figure}
	\centering
	\includegraphics[width=0.9\linewidth]{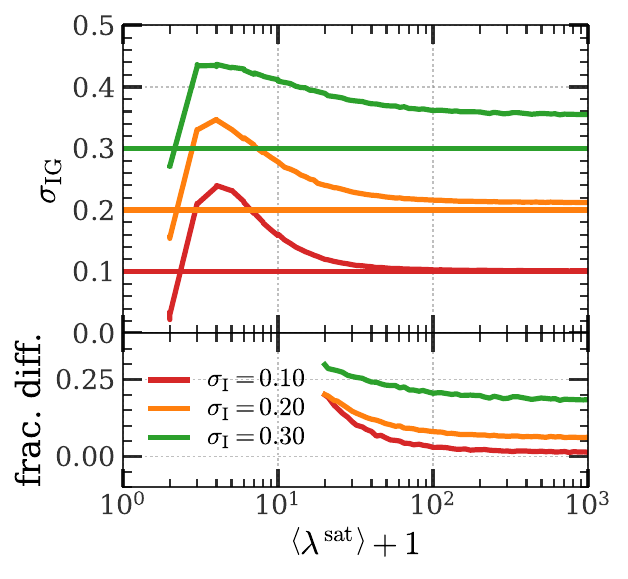}
	\caption{Richness (mass) dependence of $\sigma_\text{IG}$ derived from sampling. Each colors represents a  different value of $\sigma_\text{I}$ as indicated in the legend.
	The second row illustrates the fractional difference between $\sigma_\text{IG}$ and  $\sigma_\text{I}$.}
	\label{fig:SigmaIG}
\end{figure}

On the one hand, starting from sampling, we select a set of $\{\left< \lambda^{\text {sat }} \right>,\sigma_\text{I} \}$, sample a population of $\lambda$, calculate the mean $\left< \ln \lambda \right>$ and variance $\sigma_{\ln\lambda}$ of $\ln\lambda$, and then subtract the scatter contributed by the Poisson distribution to obtain  $ \sigma_\text{IG}^2= \sigma_{\ln \lambda }^2 - \left(e^{\left< \ln \lambda \right>} -1 \right)/e^{2\left< \ln \lambda \right>}$. \autoref{fig:SigmaIG} presents the derived values of $\sigma_\text{IG}$.
Overall, $\sigma_\text{IG}$ is larger than $\sigma_\text{I}$ and exhibits a clear mass dependence. 
Even when considering only clusters with $\lambda>20$, as done in \cite{Capasso2019} \cite{Bleem2020} and \cite{Costanzi2021}, a weak mass dependence still remains. Neglecting this dependence would lead to an overestimation of $\sigma_\text{IG}$. In the subsequent section, for the purpose of comparison with the existing literature, we choose to ignore the mass dependence of $\sigma_\text{IG}$.

\begin{figure}
	\centering
	\includegraphics[width=\linewidth]{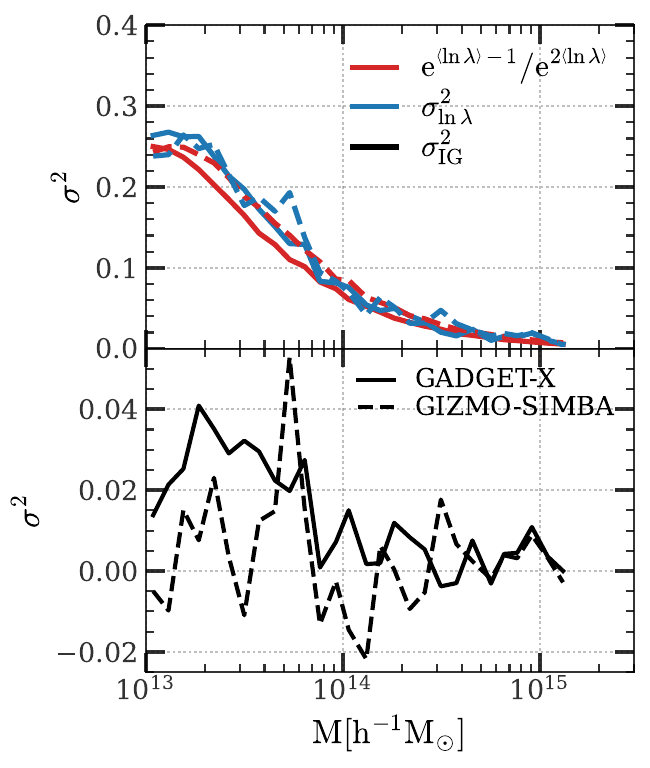}
	\caption{ Richness (mass) dependence of $\sigma_\text{IG}$ derived from simulation data at $z=0$. 
		The upper panel shows the Poisson term (red)  and the variance (blue). 
		The lower panel presents the derived intrinsic scatter (black) in the log-normal distribution.
		Solid lines correspond to the \gadgetx\ simulation, while dashed lines correspond to the \simba\ simulation.}
	\label{fig:DataSigma}
\end{figure}

On the other hand, starting from the simulation data we divide clusters into several mass bins, calculate $\left< \ln \lambda \right>$ and $\sigma_{\ln\lambda}$ in each bin, and then estimate the Poisson term and $\sigma_\text{IG}$ as shown in \autoref{fig:DataSigma}.
\autoref{fig:DataSigma} indicates a significant mass dependence of $\sigma_\text{IG}$ for \gadgetx, which is similar to \autoref{fig:SigmaIG}. While $\sigma_\text{IG}^ 2$ for \simba\ fluctuates around $0$, implying that the richness in \simba\ closely follows a Poisson distribution.

In summary, the skewed Gaussian distribution outperforms the log-normal distribution even without accounting for mass dependence. Additionally, the scatter of the log-normal distribution $\sigma_\text{IG}$ exhibits a nonlinear mass dependence, and neglecting this dependence would lead to an overestimation of the scatter. Therefore, we opt to model using the skewed Gaussian function with a scatter $\sigma_\text{I}$. At last, the same distribution function is applied to both $M_\star$ and Magnitude limits. As shown in \autoref{sec:app1}, this skewed Gaussian function also provides a good fit to the data with magnitude limit in \autoref{fig:DataDistribution4}.

\subsection{fitting procedure}\label{sec32}

 We define the richness $\lambda$ as the count of member galaxies satisfying specific selection thresholds within a halo of radius $R_{200c}$. We consider two kinds of thresholds for member selection: (1) galaxy stellar mass $M_{\star}$, and (2) galaxy absolute magnitude in the CSST i-band $ \mathscr{M}_{i}$.
 
 For each redshift and galaxy selection, we set distinct halo mass limits $M_{\text{limit}}$ that ensure the fraction of halos with a richness less than 10 $f_{\lambda<10}$ remains below 0.1 within each halo mass bin.
 We adopt this criteria for two primary reasons: (1) The corresponding $M_{\text{limit}}$ value is approximately $5\times 10^{13}-6\times 10^{14} \hMsun $, which aligns with the typical mass of a cluster $\sim 10^{14} \hMsun$, and (2) a richness below 10 leads to deviations from a power-law form of scaling relation.
 
\begin{figure}
	\centering
	\includegraphics[width=\linewidth]{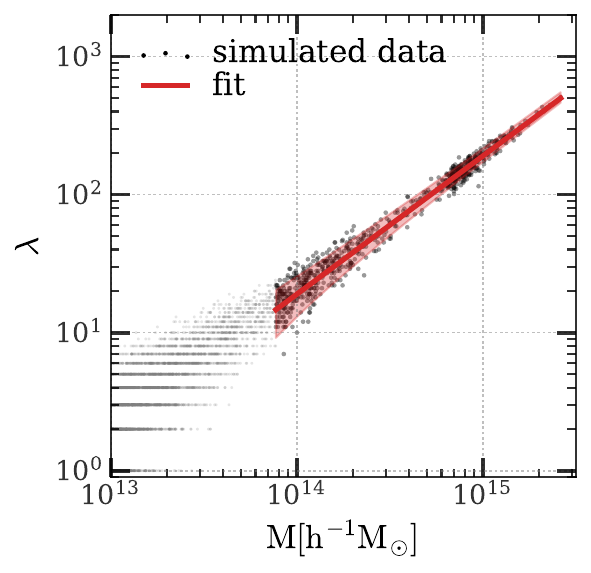}
	\caption{Mass-richness relation for \gadgetx\ at $M_{\star}=10^{9.5} \hMsun$ and $z=0$. Each dot represents an individual halo. Smaller gray dots that do not satisfy $f_{\lambda<10}<0.1$ have been discarded. The red line represents the mean relation through the fitting of \autoref{eq:lnRichness}, and the shaded area shows the 68\% confidence region of the skewed Gaussian distribution of \autoref{eq:Probability}.}
	\label{fig:Mstar_MRS_single}
\end{figure}

 To estimate parameters $\{A, B,\sigma_\text{I}\}$, we fit to the data simultaneously using the Python package \textit{emcee}, a Markov Chain Monte Carlo (MCMC) ensemble sampler developed by \cite{emcee}. In \autoref{fig:Mstar_MRS_single}, we show an example of the MR relation for \gadgetx\ with $M_{\star} \geq 10^{9.5} \hMsun$ at $z=0$. The data points are coming from the simulation and the red line and shaded region are the fitting results.

Note that for larger $M_{\star}$, not all redshifts have fitting results. This is due to the requirement on $d\log M$, the logarithmic halo mass difference between the largest halo mass and the halo mass limit $M_{\text{limit}}$, which has to be greater than 0.5. Below this value, there will not be sufficient data to constrain the slope parameter $B$. This plot confirms our fitting is working as expected, especially for the error estimation.

We have considered the mass dependence of $\sigma_\text{I}$ and found it to be consistent with 0.
 Specifically, we model $\sigma_\text{I}$ as $\sigma_\text{I}=\sigma_\text{I0}+q\times \ln (M/M_{piv})$, then fitted these four parameters $\{A, B,\sigma_\text{I0},q\}$ and finally found $q\simeq0$. So for brevity, we only consider three parameters $\{A, B,\sigma_\text{I}\}$ hereafter.
 Furthermore, we do not parameterize the redshift evolution of these parameters directly. Instead, we infer it from different redshifts $z=[0,0.5,1,1.5]$ and then examine their evolution by determining the most suitable value of a posterior, which will be detailed later.

\section{results}\label{sec04}
\begin{figure*}
	\centering
	\includegraphics[width=\linewidth]{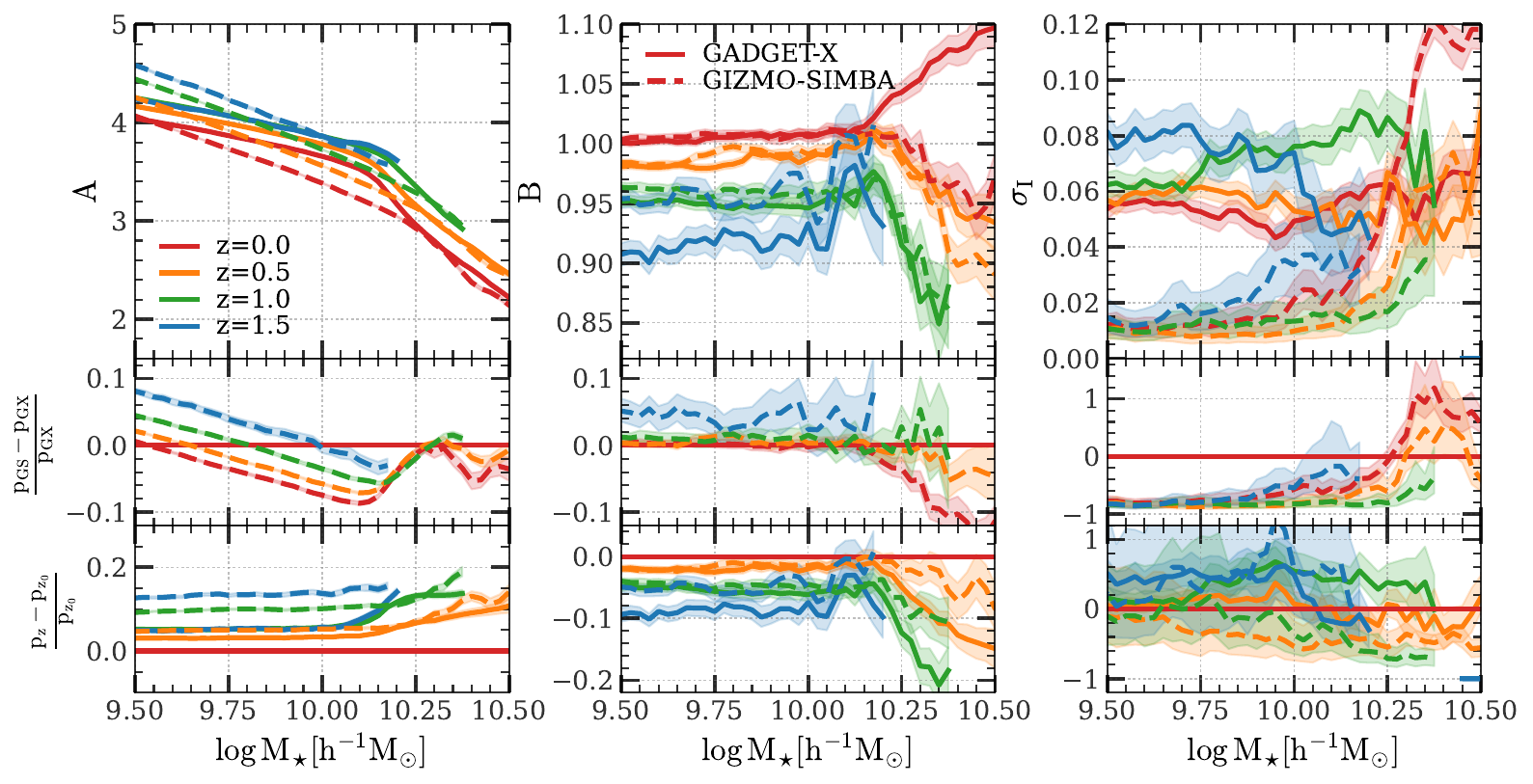}
	\caption{Fitting parameters $\{A,B,\sigma_\text{I}\}$ for the richness-mass distribution as functions of the galaxy stellar mass threshold $M_\star$, derived from the  \gadgetx\ (solid lines) and the \simba\ simulations (dashed lines).
	From left to right, the columns show $A$, $B$, and $\sigma_\text{I}$, respectively, and different colors represent different redshifts as indicated in the legend. 
	Error bands represent the 68\% confidence regions for the fitted parameters. 
	The second row shows the fractional difference between the two simulations.
	The third row illustrates the fractional difference between different redshifts for each simulation.
     }
	\label{fig:params}
\end{figure*}

In this section, we present our main results on the MR relation based on the \theth\ cluster simulations. The richness can be measured with both stellar mass and magnitude limits on member galaxies. We present the two cases separately in the following two subsections. With our fitting method described in the previous section, we only show the results of fitting parameters in this section.

\subsection{ MR relation with galaxy selection by stellar mass}\label{sec41}

For the richness based on galaxy selection by stellar mass, we adopt the galaxy stellar mass threshold $M_{\star}$ ranging from $ 10^{9.5} \hMsun$ to $10^{10.5} \hMsun$. 
The lower limit, $10^{9.5} \hMsun$, is determined by the simulation resolution \citep[see][]{Jimenez2023}. Considering that current survey can already observe galaxies with a stellar mass of $10^{10}\hMsun$ \citep{Murata2019}, our results with $M_{\star}$ in the range of $ 10^{9.5}-10^{10} \hMsun$ will be informative for future surveys.
For the upper limit, $10^{10.5} \hMsun$, we take a look at the satellite galaxy stellar mass function (SSMF). Figure 8 in \cite{Cui2022} illustrates an unrealistic peak at $10^{10.3} \hMsun$ in the SSMF for \gadgetx\ compared to the SDSS result \citep{Yang2018}, and a sharp decline around $10^{10.4} \hMsun$ for \simba, which is attributed to the AGN feedback treatment. Therefore, our results with $M_{\star}$ in the range of $ 10^{10}-10^{10.5} \hMsun$ allow for a comparison of effects within this interval, which can further identify their influences on the MR relation. Above the stellar mass upper limit, we will have only a limited number of galaxies even in clusters, which will the fitting as described in the previous section.

In \autoref{fig:params}, we present our main results on the fitting parameters $\{A, B, \sigma_\text{I}\}$ as a function of the stellar mass threshold at different redshifts depicted in different colors. Results from \gadgetx\ are presented with solid lines, while \simba\ with dashed lines. Shaded regions are the 68\% confidence intervals. The relative differences between the two simulations and different redshifts are highlighted in the middle and bottom rows, respectively.

\bigskip

The amplitude $A$ decreases with the stellar mass threshold for both simulations, which is expected. This is simply because the richness decreases as a higher stellar mass cut is applied. 
When $M_{\star} \lesssim 10^{10} \hMsun$, we demonstrate that $A$ is linearly correlated with $\log M_{\star}$. While its redshift evolution can be modeled by constant values albeit the two different simulations exhibit different evolution trends and strengths, as illustrated in the middle- and lower-left panels of \autoref{fig:params}. The constant shift indicates that there is almost no redshift evolution in the shape of the SSMF \citep{Xu2022} below $M_{\star} \lesssim 10^{10} \hMsun$. The amplitude $A$ increases with redshift, which is in line with HOD results, and is mainly due to the process of hierarchical accretion \citep{Kravtsov2004,Zheng2005,Contreras2017,Contreras2023}, see also \autoref{sec02} for more discussions on why $A$ increases with redshift. We only note here that \simba\ exhibits a larger value of $A$ at high redshifts and a smaller value at low redshifts, which can be attributed to early star formation and strong AGN feedback \citep{Cui2022};
While for \gadgetx, $A$ remains relatively constant at high redshifts.

However, when $M_{\star} \gtrsim 10^{10.25} \hMsun$, this behavior starts to be altered -- the agreement between the two simulations is much better at all redshifts; while the redshift evolution depends on the galaxy stellar mass threshold with a tilt-up. This implies a redshift evolution in the shape of SSMF in this stellar mass range. This is in agreement with the CSSMF shown in \autoref{fig:cumuSSMF}: for \gadgetx, the knee point changes from 10.1 to 10.2 when the redshift changes from 0 to 1.5. A similar behavior exists in \simba.

\bigskip

By looking at the top-central panel, the slope $B$ remains almost constant for both simulations when $M_{\star} \lesssim 10^{10} \hMsun$. Except for $z=1.5$, the agreements between the two simulations are also very good. This can also be attributed to the curve of the CSSMF which only scales with the halo mass and shows weak dependence on redshift \citep{Ahad2021}. 
The slight discrepancy between the two simulations at $z=1.5$ can be attributed to the influence of $M_{\text{limit}}$. Since there are fewer large halos at high redshift, the slope $B$ is more susceptible to $M_{\text{limit}}$. We have checked that increasing $M_{\text{limit}}$ yielded greater consistency in the values of $B$ between the two simulations at $z=1.5$.
As illustrated in the third row of \autoref{fig:cumuSSMF}, before reaching $10^{10}\hMsun$, the difference between different halo mass bins remains constant with respect to $M_\star$, and this consistency is observed in both \gadgetx\ and \simba\ simulations, which explains the agreement of $B$. However, after surpassing $10^{10}\hMsun$, the $B$ values increase for \gadgetx\ at $z=0$ and both simulations at $z=1.5$, while its values decrease for the others. Therefore, the good agreement between the two simulations still exists except for $z=0$. The reason can be explained as there are more galaxies in \simba\ than \gadgetx\ for lower halo mass, but less for higher halo mass at $z=0$ as illustrated in \autoref{fig:cumuSSMF}. While the difference between the two simulations is more or less consistent at other redshifts, i.e. \simba\ tends to have more galaxies in halos than \gadgetx\ with different masses. At last, the redshift evolution of $B$ is also constant with $M_{\star} \lesssim 10^{10} \hMsun$ and these constant values are also similar between the two simulations except for the highest redshift.

\bigskip

For \gadgetx\ over the entire range of $M_\star$ range, and for \simba\ at $M_{\star} \gtrsim 10^{10.3} \hMsun$, the scatter $\sigma_\text{I}$ remains relatively constant with $M_\star$ and similar between the two simulations. However, at $M_{\star} \lesssim 10^{10} \hMsun$, \simba\ has a much lower $\sigma_\text{I}$ compared to \gadgetx. Because this intrinsic scatter is dominated by the low-mass halos (see \autoref{fig:Mstar_MRS_single}), we think the richness in \simba\ tends to have a smaller scatter at low mass halos than \gadgetx. Though the intrinsic scatter in \simba\ shows weak dependence on stellar mass, the one in \gadgetx\ tends to present a weak increase with redshift rather than dependence on stellar mass.
Taken together, these three dependencies collectively suggest that the intrinsic scatter is likely attributed to environmental factors \citep{Mao2015}.

For \gadgetx, $\sigma_\text{I}$ demonstrates an increasing trend with redshift.
For \simba, when $M_{\star} \lesssim 10^{10} \hMsun$, $\sigma_\text{I}$ remains below 0.02 at the 68\% confidence level for all the redshifts, consistent with \autoref{fig:DataSigma}. This suggests that the richness in \simba\ follows a nearly Poisson distribution, even at large occupation numbers. This behavior can be attributed to the intense baryonic processes in \simba, resulting in a negligible environmental impact relative to the strength of the baryonic processes.
However, when $M_{\star} \gtrsim 10^{10}\hMsun$, $\sigma_\text{I}$ increases rapidly and shows a decreasing trend with redshift up to $z=1$, which is opposite to the \gadgetx\ run.

In summary, when $M_{\star} \gtrsim 10^{10} \hMsun$, the behavior of parameters displays stronger influence by the baryon models. When $M_{\star} \lesssim 10^{10} \hMsun$, the dependence of our parameters, $A$ and $B$, on redshift and stellar mass, is consistent with certain findings of the HOD studies at large \citep{Kravtsov2004,Zheng2005,Contreras2017,Contreras2023}. However, comparing our results to \cite{Contreras2017} and \cite{Contreras2023}, there exist subtle differences in the redshift dependence. Specifically, our parameter $A$ for \gadgetx\ remains roughly to be a constant at higher redshift, whereas their $A$ demonstrates an increase with $z$ which agrees better with \simba. Moreover, the slope $B$ from observations remains a constant for redshifts greater than approximately $0.7$, while our $B$ shows a decreasing trend for both simulations. 
These distinctions could be attributed to different galaxy selections. In contrast to their approach of fixing the galaxy number density $n$ for different redshifts, we maintain a fixed galaxy stellar mass threshold $M_\star$. We have checked that if we fix $n$,  $A$ exhibits an increasing trend with redshift. However, $B$, at least until $z=1.5$, continues to show a downward trend which indicates the richnesses for different halo masses have fewer variations.

\subsection{MR relation with galaxy selection by magnitude}\label{sec42}
\begin{figure*}
	\centering
	\includegraphics[width=\linewidth]{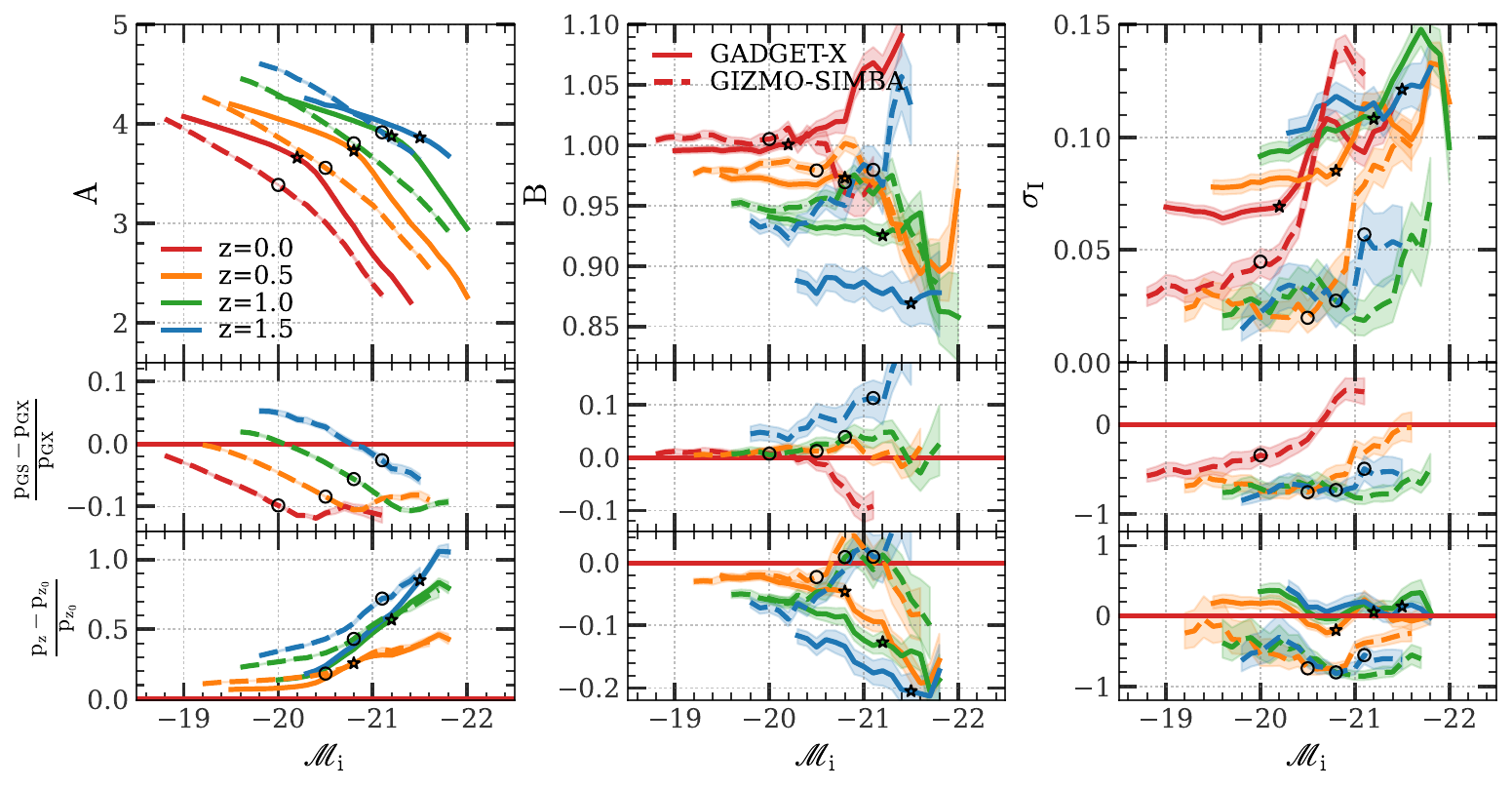}
	\caption{
      Fitting parameters $\{A,B,\sigma_\text{I}\}$ for the richness-mass distribution as functions of the threshold on galaxy's absolute magnitude in  the CSST i-band $\mathscr{M}_i$. Labels and legends are the same as \autoref{fig:params}.
    Star markers (circle markers) correspond to $M_\star =10^{10} \hMsun$ according to the $M_\star-\mathscr{M}_i$ relation (\autoref{eq:MstarMag}) for \gadgetx(\simba).}

	\label{fig:params_i}
\end{figure*}
 
Galaxy stellar mass is not a quantity that can be directly measured from observation. However, it is closely related to the galaxy's luminosity or magnitude. As such, the richness can be also derived with selection of galaxies based on their magnitudes. In this subsection, we investigate the MR relation when the galaxy magnitude limit, instead of galaxy stellar mass limit, is used for controlling the richness. 

We utilize limit on the absolute magnitude as the galaxy selection criteria, employing the CSST i-band $\mathscr{M}_i $ with $\mathscr{M}_i $ ranging from $-18$ to $-23$. The fitting results of parameters $\{A, B, \sigma_\text{I}\}$ as functions of $\mathscr{M}_i $ are depicted in \autoref{fig:params_i}, which is similar to \autoref{fig:params}.
We just show the results of $\mathscr{M}_i $ corresponding to the range of $ M_{\star}=[ 10^{9.5}, 10^{10.5} ] \hMsun$, and mark the point corresponding to $M_\star =10^{10} \hMsun$ through the $M_\star-\mathscr{M}_i$ relation as shown in \autoref{sec51}.

In general, if $\log M_\star $ is correlated with the magnitude without scatter, we would expect that the fitting parameters of the MR relation will be a simple shift from those with cuts on $M_\star$. By comparing \autoref{fig:params_i} and \autoref{fig:params}, we find the conclusions in the previous subsection are qualitatively unchanged. More discussions on the $\log M_\star $ and magnitude relation can be found in \autoref{sec51}. Here we focus on the subtle changes in the fitting parameters.

The dependence of $A$ and $B$ on redshift and galaxy threshold remains consistent with \autoref{sec41}, with $M_\star \approx 10^{10} \hMsun$ serving as the dividing point. 
However, the redshift evolution around $M_\star \approx 10^{10} \hMsun$ seems to be not consistent with the fainter galaxy end, unlike what has been shown in \autoref{fig:params}. $M_\star \approx 10^{10} \hMsun$ is more-or-less the separation part in the galaxy color bimodality plot, which contains both blue, star-forming and red, quenched galaxies. \MJ{When $M_\star \geq 10^{10} \Msun$, Fig.9 in \citet{Cui2022} exhibits a clear separation in the satellite galaxy color-magnitude diagram between \gadgetx\ and \simba\, with galaxies in \simba\ appearing blue.} We know that a galaxy's luminosity is strongly dependent on its color, as such, it is not surprising to see an increased scatter around that stellar mass, which results in an increase of $\sigma_\text{I}$ for both \gadgetx\ and \simba\ as is shown in the third column. In addition, this separation varies with redshift because of more star-forming galaxies at higher redshift. With this additional dependence, i.e. more brighter galaxies at higher redshift, the redshift evolution behaves differently from the case with $M_\star$ limits for all the three parameters. 

\section{discussions}\label{sec05}
\subsection{conversion between $M_\star$ and $\mathscr{M}_i$}\label{sec51}

\begin{figure}
\centering
\includegraphics[width=\linewidth]{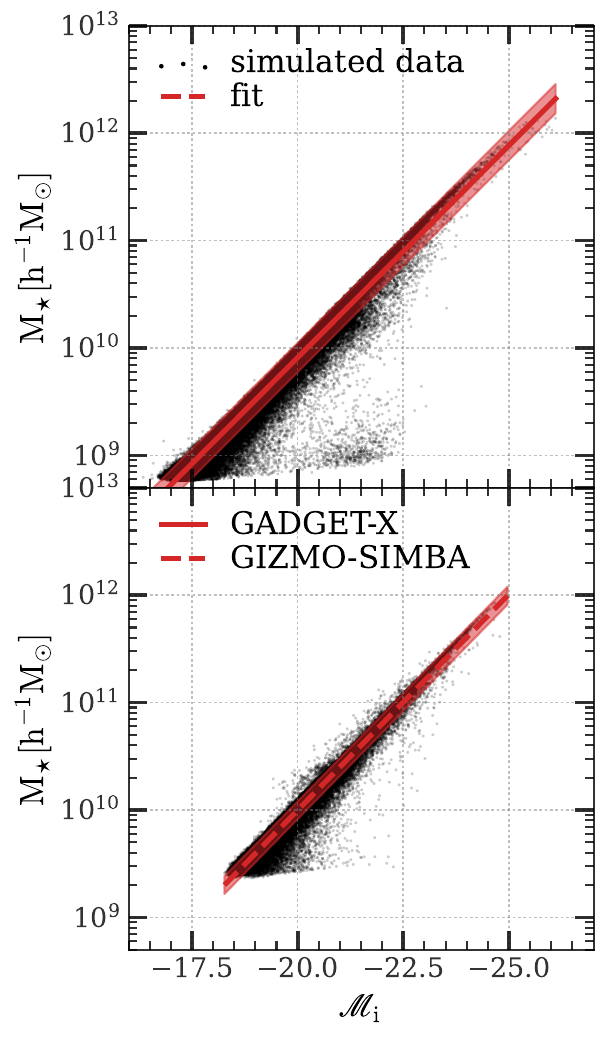}
\caption{Galaxy stellar mass-absolute magnitude $M_\star-\mathscr{M}_i$ relation for \gadgetx(the upper panel) and \simba(the lower panel) in CSST i-band at $z=0$. Each dot represents an individual galaxy.
The red line represents the mean relation Equation \eqref{eq:MstarMag}, and the contour indicates the 68\% confidence region of the Gaussian error.}
  \label{fig:mag_MmS}
\end{figure}
		
\begin{figure*}
\centering
\includegraphics[width=\linewidth]{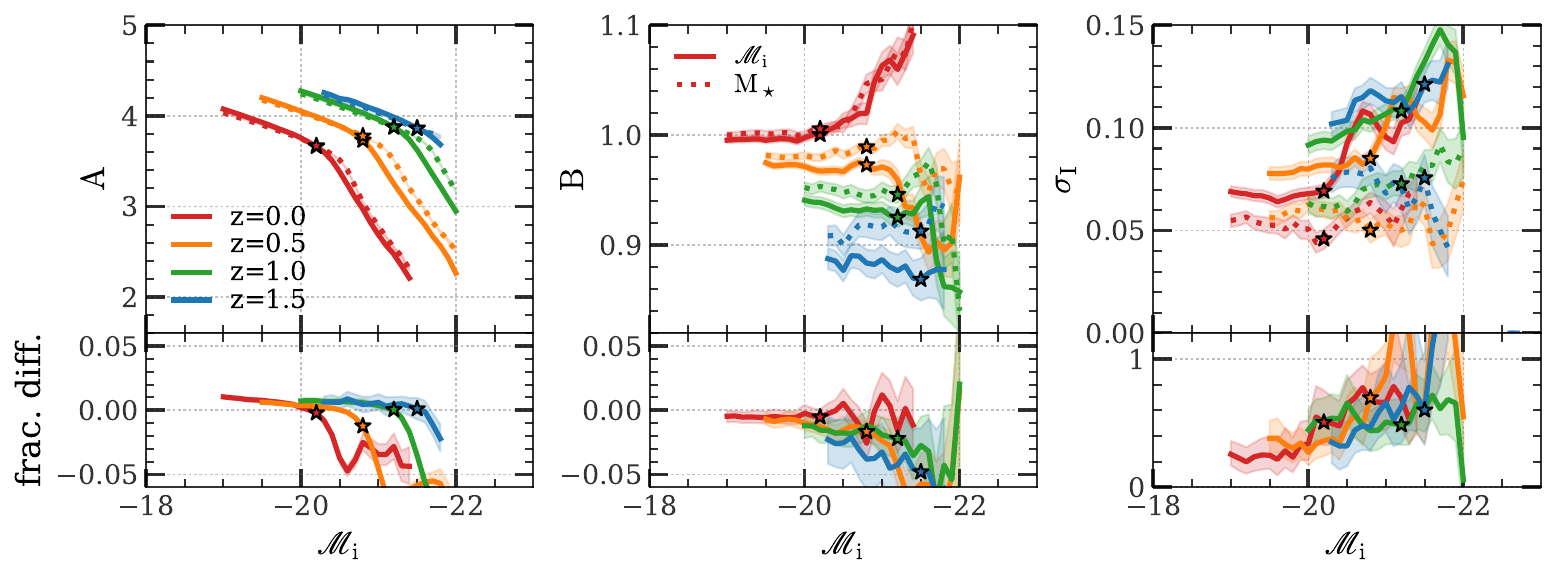}
\caption{Fitting parameters $\{A,B,\sigma_\text{I}\}$ varying with the absolute magnitude $\mathscr{M}_i$, just for \gadgetx.
	Solid lines correspond to galaxies directly selected using $\mathscr{M}_i$, while dotted lines represent galaxies selected based on $M_\star$ converted from $\mathscr{M}_i$ using Equation \eqref{eq:MstarMag}.
	From left to right, each column corresponds to the properties $A$, $B$, and $\sigma_\text{I}$, respectively, and different colors indicate different redshifts as indicated in the legend. 
	Error bars represent the standard deviation from the fitted parameter value. 
	The second row illustrates the fractional difference between these two selection methods.
    The black star points denote $M_\star=10^{10} \hMsun$.}
\label{fig:Mstar_to_mag_GadgetX_abs_csst_i}
\end{figure*}

Practical sky surveys employ the magnitude to select galaxies, rather than the stellar mass. However, it is not realistic to provide fitting results of $\{A, B, \sigma_\text{I}\}$ for all bands in all surveys. Consequently, we aim to investigate whether it is possible to derive MR relations based on different galaxy magnitudes from a single MR relation using the $M_\star$ threshold in \autoref{sec41}.
To accomplish this, we naturally turn to the galaxy stellar mass-absolute magnitude relation $M_\star-\mathscr{M}$ and specifically focus on the CSST i-band $\mathscr{M}_i$ as an illustrative example.

We use a simple linear relation \MJ{$\ln M_\star = A_i + B_i \mathscr{M}_i + C_i \ln (1+z)$, along with a Gaussian probability function $P(\ln M_\star|\mathscr{M}_i)$ incorporating a magnitude-redshift-independent scatter $\sigma_i$, to model the $M_\star-\mathscr{M}_i$ relation. The four parameters $\{A_i, B_i, C_i, \sigma_i$\} are simultaneously fitted using the same procedure described in \autoref{sec32}, but with galaxies as the input data. The resulting $M_\star-\mathscr{M}_i$ relations, without showing $\sigma_i$,} for \gadgetx\ and \simba\ \MJ{are as follows, }respectively:
		\begin{eqnarray}\label{eq:MstarMag}
\ln M_\star=4.63 -0.91  \mathscr{M}_i -1.30  \times  \ln (1+z), \nonumber
	\\
	\ln M_\star=4.49 -0.93  \mathscr{M}_i -1.10  \times  \ln (1+z).
\end{eqnarray}	
An example has been shown in \autoref{fig:mag_MmS} at $z=0$.
By employing this relation, we convert $\mathscr{M}_i$ into $M_\star$ as the selection criteria. The corresponding results are depicted by the dotted lines in \autoref{fig:Mstar_to_mag_GadgetX_abs_csst_i}. The solid lines, on the other hand, represent the outcomes obtained directly using $\mathscr{M}_i$.

Parameters $\{A, B\}$ obtained from these two selections exhibit consistency within a fractional difference of 5\% across the entire range, especially small with $M_\star \lesssim 10^{10} \hMsun$. It is worth noting that $B$ with magnitude limit has a consistently lower value compared to the one with $M_\star$ limit, the difference increases with redshift.
In addition, the scatter $\sigma_\text{I}$ obtained using $\mathscr{M}_i$ is significantly larger than the scatter $\sigma_\text{I}$ obtained using $M_\star$. This difference arises due to the presence of scatter in the $M_\star-\mathscr{M}_i$ relation, which increases $\sigma_\text{I}$ by approximately 50\%. 
As indicated in the previous section, the large scatter as well as the redshift dependence of the parameters closely connect with the galaxy formation, especially the galaxy quenching event. As such, directly using the MR fitting result with magnitude cuts to estimate halo masses should be careful, an improper simulation, especially one that can not provide a faithful galaxy color-magnitude diagram at multiple redshifts, may lead to biased results.

Nevertheless, these findings based on our simulations indicate that it is feasible to derive magnitude threshold results from stellar mass threshold results by utilizing the stellar mass-magnitude $M_\star-\mathscr{M}_i$ relation.
Importantly, these conclusions are applicable not only for \gadgetx\ in CSST-i band, but also in other bands, as well as for \simba. A comprehensive presentation of these results is provided in the appendix. 

\MJ{It is noteworthy that the $M_\star-\mathscr{M}_i$ relation fitted from the simulation is based on the FSPS code, in which we select the initial mass function (IMF) of \cite{Chabrier2003}, consistent with both simulations' setups. Adopting different IMFs will change the galaxy's magnitude \citep[e.g.][]{Cappellari2012, Narayanan2012, Bernardi2018}. Generally speaking, the top-heavy IMF is found in regions of elevated star formation rate \citep[e.g.][]{Gunawardhana2011}, which will yield more light in high energy bands.}

\subsection{7-parameters relation}

\begin{figure}
	\centering
	\includegraphics[width=0.9\linewidth]{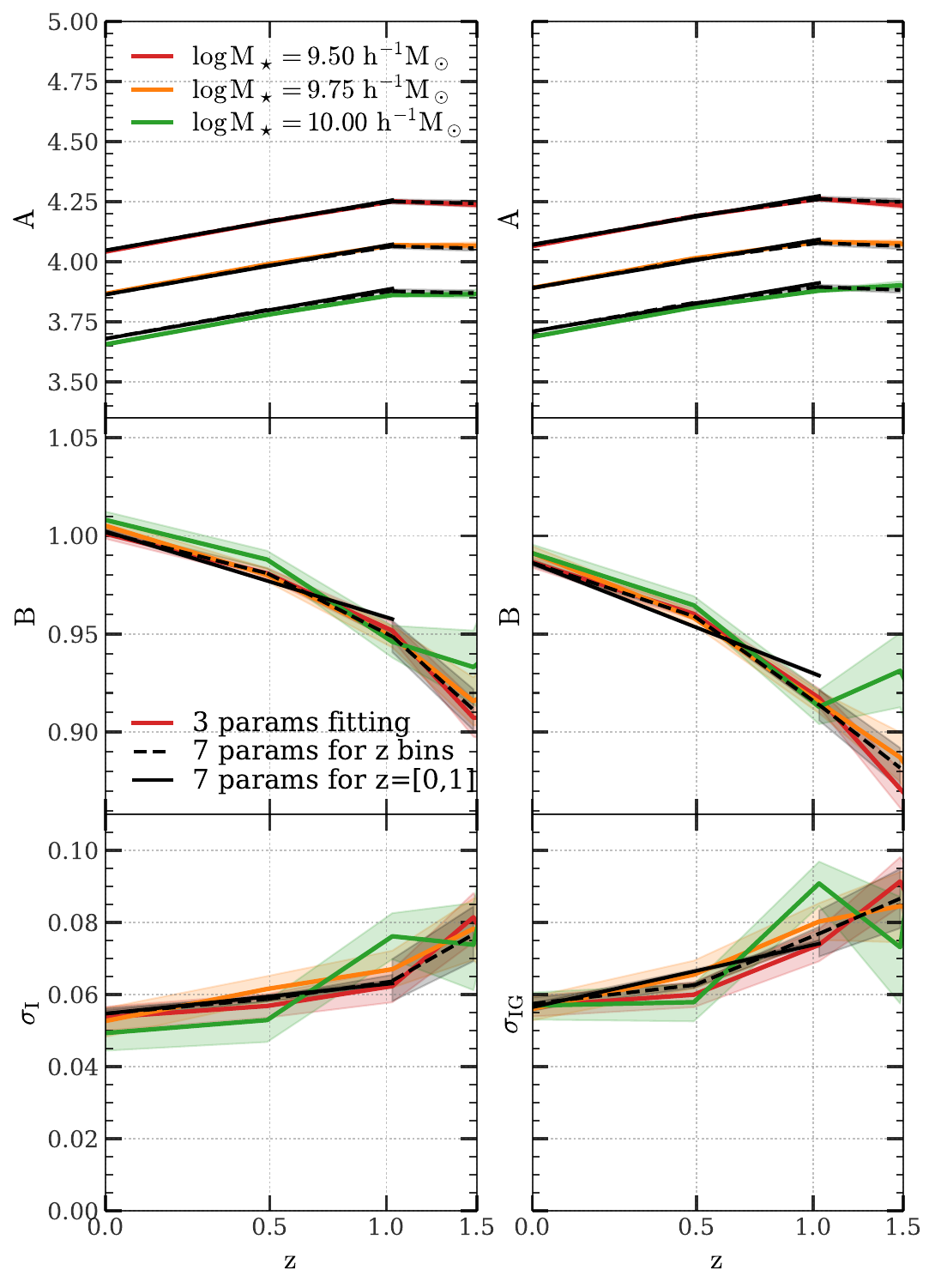}
	\caption{Fitting results for \gadgetx.
		The left panel displays the results obtained using the skewed Gaussian distribution, while the right panel shows the results obtained using the log-normal distribution.
		Colored lines represent the 3-parameters fitting performed at specific redshifts and stellar mass thresholds, as done in \autoref{sec04}.
		Black lines represent the 7-parameters fitting conducted over a range of redshifts(dashed lines for $z=[0,0.5],\ [0.5,1]$ and $[1,1.5]$; the solid line for $z=[0,1]$) and stellar mass thresholds, as described in this section.
	}
	\label{fig:Mstar_paramsZGadgetX}
\end{figure}

\begin{figure}
	\centering
	\includegraphics[width=0.9\linewidth]{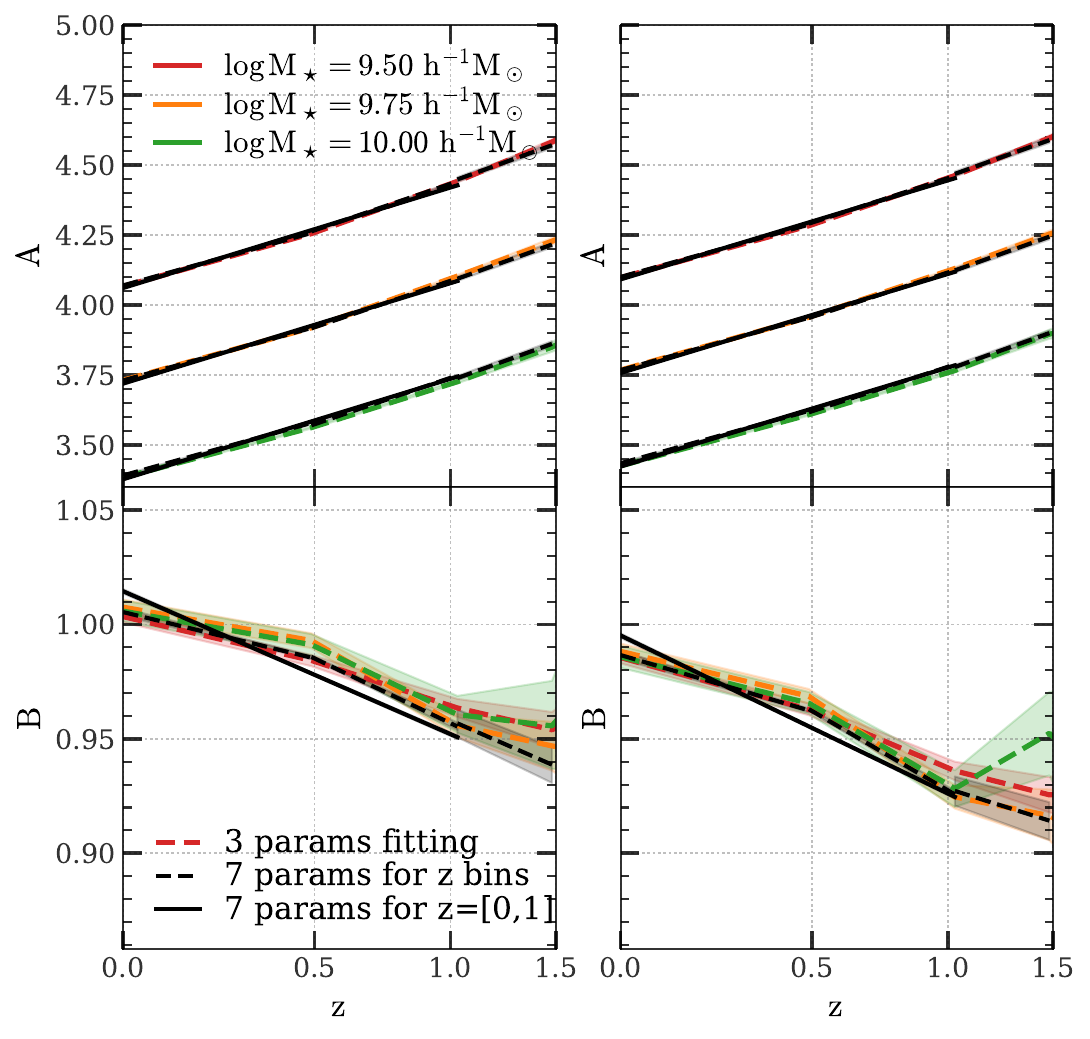}
	\caption{Similar to \autoref{fig:Mstar_paramsZGadgetX}, but for \simba. Scatter parameters have been omitted either.}
	\label{fig:Mstar_paramsZSIMBA}
\end{figure}

From this section, we focus only on the range of $ M_\star= 10^{9.5}-10^{10} \hMsun$, considering the current and future survey limits and the clearer stellar mass trends before $10^{9.5} \hMsun$ in the \autoref{fig:params}.

We consider two distributions as mentioned previously, a skewed Gaussian distribution $P(\lambda)$ (Equation \eqref{eq:Probability}) with the scatter $\sigma_\text{I}$, and a log-normal distribution $P(\ln \lambda)$ with the scatter $\sigma_\text{IG}$. We refer interesting readers to \autoref{sec:app2} for detailed comparisons. Despite the small scatter for \simba, we still utilize both distributions because different distributions have an impact on the fitting of parameters $A$ and $B$.

In \autoref{sec41}, we have presented the redshift and stellar mass dependencies. Now, we incorporate both dependencies into the calculation:
\begin{eqnarray}\label{eq:7params}
	A \rightarrow A_0+A_z \times \ln \frac{1+z}{1+z_p} +A_\star \times \ln \frac{M_\star}{M_{\star p}}, \nonumber
	\\
	B \rightarrow B_0+B_z \times \ln \frac{1+z}{1+z_p},  \nonumber
	\\
	\sigma_\text{I} \rightarrow \sigma_\text{I0}+\sigma_z \times \ln \frac{1+z}{1+z_p},
\end{eqnarray}
for the skewed Gaussian distribution, and: 
\begin{eqnarray}\label{eq:7params1}
	A \rightarrow A_0+A_z \times \ln \frac{1+z}{1+z_p} +A_\star \times \ln \frac{M_\star}{M_{\star p}}, \nonumber
	\\
	B \rightarrow B_0+B_z \times \ln \frac{1+z}{1+z_p},  \nonumber
	\\
	\sigma_\text{IG} \rightarrow \sigma_\text{IG0}+\sigma_z \times \ln \frac{1+z}{1+z_p},
\end{eqnarray}
for the log-normal distribution, where $z_p=0.5$, $M_{\star p}=1\times 10^{10} \hMsun$.

Now there are a total of 7 parameters, namely $\{A_0$, $A_z$, $A_\star$, $B_0$, $B_z$, $\sigma_\text{I0}, \sigma_z\}$ or $\{A_0$, $A_z$, $A_\star$, $B_0$, $B_z$, $\sigma_\text{IG0}$, $\sigma_z\}$. These 7 parameters replace the 3 parameters used previously, and we repeat the fitting procedure for all clusters at redshifts $z=[0, 1.5]$ with a redshift interval of $\mathrm{d}z=0.5$. We set galaxy stellar mass thresholds of $M_\star=[10^{9.5}, 10^{10}] \hMsun$ with a mass interval of $\mathrm{d}\log M_\star = 0.025$. To better capture the redshift evolution, we perform piecewise fitting for different redshift intervals. 
The results of these fits are presented in \autoref{tab:Mstar_paramsZGadgetX} and \autoref{fig:Mstar_paramsZGadgetX} for \gadgetx, and \autoref{tab:Mstar_paramsZSIMBA} and \autoref{fig:Mstar_paramsZSIMBA} for \simba.

\begin{table}
	\caption{\label{tab:Mstar_paramsZGadgetX}
		The 7 fitting parameters for \gadgetx. The upper panel displays the results obtained using the skewed Gaussian distribution, while the lower panel shows the results obtained using the log-normal distribution. Each column corresponds to a different redshift range. Fitting errors smaller than 10\% have been omitted for a cleaner presentation.}
	\begin{ruledtabular}
		\begin{tabular}{l|c|ccc} 
			$z$ &[0,1]&[0,0.5]&[0.5,1]&[1,1.5] \\
			\hline
			$A_0$& 3.792& 3.803& 3.800& 3.887\\
			$A_z$& 0.205& 0.245& 0.150& $-0.017 _{- 0.006 }^{+ 0.006 }$\\
			$A_\star$& -0.320& -0.319& -0.323& -0.325\\
			$B_0$& 0.980& 0.981& 0.980& 0.993\\
			$B_z$& -0.031& -0.042& -0.060& -0.083\\
			$\sigma_\text{I0}$& 0.060& 0.059& 0.059& 0.048\\
			$\sigma_z$& $0.008 _{- 0.001 }^{+ 0.001 }$& $0.008 _{- 0.002 }^{+ 0.002 }$& $0.009 _{- 0.002 }^{+ 0.002 }$& $0.029 _{- 0.004 }^{+ 0.004 }$\\
			\hline
			$A_0$& 3.819& 3.833& 3.829& 3.911\\
			$A_z$& 0.196& 0.244& 0.128& $-0.028 _{- 0.007 }^{+ 0.006 }$\\
			$A_\star$& -0.314& -0.313& -0.316& -0.318\\
			$B_0$& 0.957& 0.959& 0.958& 0.952\\
			$B_z$& -0.044& -0.056& -0.084& -0.072\\
			$\sigma_\text{IG0}$& 0.067& 0.063& 0.063& 0.066\\
			$\sigma_z$& 0.019& $0.011 _{- 0.002 }^{+ 0.002 }$& 0.026& $0.021 _{- 0.004 }^{+ 0.005 }$\\
		\end{tabular}
	\end{ruledtabular}
\end{table}

\begin{table}
	\caption{\label{tab:Mstar_paramsZSIMBA}
		Similar to \autoref{tab:Mstar_paramsZGadgetX}, but for \simba. Scatter parameters have been omitted because of the incomplete posterior distribution.}
	\begin{ruledtabular}
		\begin{tabular}{l|c|ccc} 
			$z$ &[0,1]&[0,0.5]&[0.5,1]&[1,1.5] \\
			\hline
			$A_0$& 3.575& 3.584& 3.575& 3.596\\
			$A_z$& 0.360& 0.383& 0.330& 0.271\\
			$A_\star$& -0.594& -0.589& -0.602& -0.616\\
			$B_0$& 0.984& 0.985& 0.985& 0.978\\
			$B_z$& -0.037& -0.041& -0.057& $-0.040 _{- 0.004 }^{+ 0.004 }$\\
			\hline
			$A_0$& 3.617& 3.627& 3.621& 3.638\\
			$A_z$& 0.353& 0.382& 0.316& 0.268\\
			$A_\star$& -0.581& -0.578& -0.585& -0.599\\
			$B_0$& 0.960& 0.962& 0.962& 0.943\\
			$B_z$& -0.043& -0.048& -0.069& $-0.030 _{- 0.004 }^{+ 0.005 }$\\
		\end{tabular}
	\end{ruledtabular}
\end{table}

In general, there are almost neglectable differences for both $A$ and $B$ fitting parameters between 3- and 7-parameter fitting. $\sigma_\text{I}$ shows a slightly larger between the two fittings. Nevertheless, the largest increase from $\log M_\star = 10 \hMsun$ is still within 0.02, which could be caused by the sample difference.

Compared to the scatter $\sigma_\text{I}$ fitted by the skewed Gaussian distribution, the scatter $\sigma_\text{IG}$ fitted by the log-normal distribution is larger as we expect.  Additionally, the log-normal distribution tends to produce larger values for the amplitude $A$ and smaller values for the slope $B$.

$A_\star$ remains constant across all redshift ranges. This indicates that there is no dependence between redshift and the stellar mass dependence of $A$, as we illustrated before. 
On the other hand, $A_z$ and $B_z$ exhibit slight variations among different redshift ranges, particularly for \gadgetx\ at $z=[1,1.5]$. This suggests that a linear fit of the redshift may not be the optimal choice, but for the subsequent comparison, a linear fit of $z=[0,1]$ is still employed.

\subsection{comparison with previous work}\label{sec53}

\begin{figure*}
	\centering
	\includegraphics[width=0.9\linewidth]{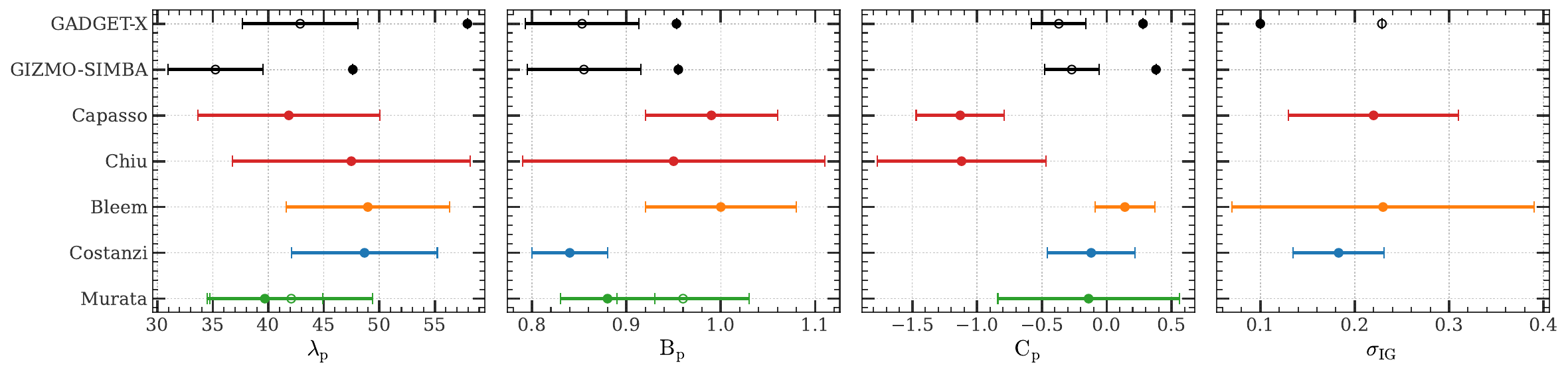}
	\caption{Parameters listed in \autoref{tab:papers}. 
		Colored dots represent parameters derived from the literature.
		Black dots show our results for full galaxies. 
		Gray hollow dots indicate that the red fraction from \cite{Hennig2017} is taken into account.
		Green hollow dots represent results in the middle redshift range in Table 3 of \cite{Murata2019}.
	}
	\label{fig:ComparePapers_params}
\end{figure*}

In this section, we present a comparative analysis of our results with various forward-modeling studies conducted by different surveys.

\cite{Capasso2019} and \cite{Chiu2023} utilize a cluster sample selected by X-ray and confirmed by optical data. The former study uses galaxy dynamical information while the latter uses cluster abundance (referred to as number counts, NC) to calibrate the MR relation.
\cite{Bleem2020} has a similar approach to \cite{Chiu2023}, but instead of X-ray, they utilize the SZ effect.
Additionally, \cite{Costanzi2021} incorporates other observable-mass relations (OMR) to supplement the information.

These studies adopt different mass definitions and relation forms. 
To facilitate comparison, we convert their respective cluster mass definitions to $M=M_{200c}$ by assuming a Navarro, Frenk, and White (NFW) profile \citep{NFW1997} and employing the concentration-mass relation from \cite{Duffy2008}. 
We then calculate the richness, as well as the dependence of richness on mass and redshift around the pivot point $M_p=3\times 10^{14}\hMsun, z_p=0.5$, based on the different redshift and richness ranges reported in the literature.
More specifically, we define:
\begin{eqnarray}\label{eq:lambdaBCp}
    &&\lambda_p \equiv \lambda(M_p,z_p), 
    \nonumber \\
	&&B_p \equiv \frac{\ln \lambda( M_2,z_p)-\ln \lambda( M_1,z_p)}{2\Delta_{\ln M}},
    \nonumber \\
    &&C_p \equiv \frac{\ln \lambda( M_p,z_2)-\ln \lambda( M_p,z_1)}{2 \Delta_{\ln(1+z)}},
\end{eqnarray}
with small enough steps $\Delta_{\ln M}=\ln\frac{M_2}{M_p }=\ln\frac{M_p}{M_1 }=0.001$ and $\Delta_{\ln(1+z)}=\ln\frac{(1+z_2)}{(1+z_p) }=\ln\frac{(1+z_p)}{(1+z_1) }=0.001$.

A summary of the aforementioned papers, along with the derived parameters, are presented in \autoref{tab:papers} and \autoref{fig:ComparePapers_params}.

Our comparison results start with an absolute magnitude threshold $\mathscr M_i=-19.47$ at $z=0$, corresponding to 0.2 times the characteristic luminosity $0.2 L_*$ applied in the redMaPPer algorithm \citep{Rykoff2012,Rykoff2014}.
The threshold varies with redshift due to the passive evolution of the stellar population. To calculate the evolution and determine the threshold at the pivot redshift $z_p$, we utilize the FSPS code. More specifically, in the evolution model, we assume that the stellar population was formed at a redshift of $z_f = 3$, and adopt the 'MIST' stellar isochrone libraries \citep{Choi2016}, the 'MILES' stellar spectral libraries \citep{Vazdekis2010}, \MJ{the IMF of \citet{{Chabrier2003}}} and the Solar metallicity.
Ultimately, we obtain the threshold at $z_p$ as $\mathscr M_i=\MJ{-19.98}$. 

Next, we employ Equation \eqref{eq:MstarMag} to obtain $M_\star$, which is $\MJ{4.77}\times 10^{9}\hMsun$ for \gadgetx, and $\MJ{6.70}\times 10^{9}\hMsun$ for \simba. Subsequently, by applying 7-parameters fitting results based on a log-normal distribution at redshifts $z=[0,1]$, the upper panel and the first column from \autoref{tab:Mstar_paramsZGadgetX} and \autoref{tab:Mstar_paramsZSIMBA}, we obtain the MR relation. 
The first two rows of \autoref{tab:papers} present this relation in the form of $\{\lambda_p, B_p, C_p\}$ using Equation \eqref{eq:lambdaBCp} for convenient comparison with others papers.
Note that the scatter here is the result of multiplying by 1.5, which is due to the transition from threshold $M_\star$ to $\mathscr{M}_i$ in \autoref{sec42}.

Furthermore, it is important to note that the cluster finders used in the referenced papers only identify red-sequence galaxies, whereas our analysis does not distinguish between red and blue galaxies. So we incorporate the red sequence fraction $f_{RS}$ Equation (13) from \cite{Hennig2017} \MJ{:
\begin{equation}
    f_{RS}(M,z)=A_{RS}
    \left(\frac{M}{6\times 10^{14}M_\sun}\right)^{B_{RS}}
    \left(\frac{1+z}{1+0.46}\right)^{C_{RS}}.
\end{equation}
Using the galaxy population of 74 SZ effect selected clusters from the SPT survey, \cite{Hennig2017} obtain $A_{RS}=0.68\pm0.03$, $B_{RS}=-0.10\pm 0.06$ and $C_{RS}=-0.65\pm0.21$. Converting this relation from their pivot point $(M=6\times 10^{14}M_\sun,z=0.46)$ to our pivot point $(M_p=3\times 10^{14}\hMsun, z_p=0.5)$ only affects the normalization parameter, resulting in $A_{RS}=0.74\pm 0.09$.
Finally, the richness with red galaxies is represented as $\ln \lambda^\text{red} = \ln \lambda +\ln f_{RS}$, and the corresponding parameters $\{\lambda_p, B_p, C_p\}$} are shown in rows 3,4 of \autoref{tab:papers}.

\begin{table*}
	\caption{\label{tab:papers}
		Richness–mass–redshift relation parameters from this analysis and the literature. 
		$\lambda_p$ is richness at the pivot point $M_p=3\times 10^{14}\hMsun, z_p=0.5$. 
		$B_p$ and $C_p$ denote the mass and redshift dependencies, respectively, around the pivot point.
 }
	\begin{ruledtabular}
		\begin{tabular}{llcccc} 
			Simulations & $M_\star[\hMsun] $ at $z_p$&$\lambda_p$ & $M^{B_p}$ &$(1+z)^{C_p} $& $\sigma_\text{IG}$ \\
			\hline
			\gadgetx\ & $\MJ{4.77}\times 10^{9}$& $\MJ{57.94}$ & $0.953$ &$\MJ{0.28}$&$0.10$\\
			\simba\ & $\MJ{6.70}\times 10^{9}$& $\MJ{47.62}$ & $0.955$ &$\MJ{0.38}$&$/$\\
			\hline
			Red galaxies  & \multicolumn{5}{l}{$\ln f_{RS}=\ln(0.74\pm 0.09)-(0.10\pm 0.06)\ln\frac{M}{M_p}-(0.65\pm0.21) \ln\frac{1+z}{1+z_p}$} \\
			\hline
			\gadgetx\ & $\MJ{4.77}\times 10^{9}$& $\MJ{42.88\pm 5.21}$ & $0.853 \pm 0.06$ &$\MJ{-0.37}\pm 0.21$&$0.23$\\
			\simba\ & $\MJ{6.70}\times 10^{9}$& $\MJ{35.24\pm 4.29}$ & $ 0.855 \pm 0.06$ &$\MJ{-0.27} \pm 0.21$&$/$\\
			\hline
			Authors &Description&$\lambda_p$ & $M^{B_p}$ &$(1+z)^{C_p} $& $\sigma_\text{IG}$ \\
			\hline
			\cite{Capasso2019}& ROSAT galaxy dynamics& $41.85 _{- 8.18 }^{+ 7.98 }$ & $0.99 _{- 0.07 }^{+ 0.06 }$ & $-1.13 _{- 0.34 }^{+ 0.32 }$ &$0.22_{-0.09}^{+0.08}$\\
			\cite{Murata2019}& HSC NC & $39.69 _{- 5.22 }^{+ 5.71 }$ & $0.88 _{- 0.05 }^{+ 0.05 }$ & $-0.14 _{- 0.70 }^{+ 0.59 }$&/\\

			\cite{Bleem2020}& SPT NC & $48.97 _{- 7.36 }^{+ 8.01 }$ & $1.00 _{- 0.08 }^{+ 0.08 }$ & $0.14 _{- 0.23 }^{+ 0.23 }$ &  $0.23_{-0.16} ^{+0.16}$\\
			\cite{Costanzi2021}& DES NC + SPT OMR& $48.67 _{- 6.56 }^{+ 5.55 }$ & $0.84 _{- 0.04 }^{+ 0.04 }$ & $-0.12 _{- 0.34 }^{+ 0.34 }$&
			 $0.207_{-0.045}^{+0.061}$ \\
			\cite{Chiu2023}& eROSITA NC& $47.48 _{- 10.69 }^{+ 11.67 }$ & $0.95 _{- 0.16 }^{+ 0.17 }$ & $-1.12_{- 0.65 }^{+ 0.54 }$ &/\\	\end{tabular}
	\end{ruledtabular}
\end{table*}

Each survey has its own strategy, so variations in $\lambda_p$ are tolerable. Additionally, although most of them utilize redMaPPer as the cluster finder, the choice of filter bands differs, which can also impact the results.

Turning to $B_p$, \cite{Hennig2017} shows a decreasing mass trend of $f_{RS}$. But regardless of the inclusion of blue galaxies, mass trends of the MR relations are consistent in their work. In contrast, \cite{Okabe2019} illustrates a weakly increasing mass dependence of $f_{RS}$ in their Figure 5.
Our $B_p$ values for full galaxies (black dots in \autoref{fig:ComparePapers_params}) demonstrate better consistency with the ICM-selected samples (red and orange dots). The other two optical-selected samples (blue and green dots) yield consistent results that are slightly smaller than ours. This divergence may be attributed to projection effects, which the ICM-selected cluster sample is not susceptible to.
\cite{Murata2019} states that their results in the middle redshift range $z=[0.4, 0.7]$ are least affected by projection effects in Table 3. Specifically, they report a slope of $0.96 _{-0.07}^{+0.09}$ for $M_{200c}$ (green hollow dots in \autoref{fig:ComparePapers_params}), which helps mitigate inconsistencies.

Now turning to $C_p$, while most observations tend to indicate a negative $C_p$, our results show positive values (black dots). However, after accounting for $f_{RS}$, these values turn out to be negative (gray dots) and exhibit more consistency with optical-selected samples (blue and green dots).

Finally, shifting our focus to scatter, previous studies by \cite{Capasso2019}, \cite{Costanzi2021} and \cite{Bleem2020} have modeled scatter using the same form as Equation \eqref{eq:sigmaIG}, albeit without considering the redshift dependence. Nevertheless, their values of $\sigma_\text{IG}$ align with each other.
When accounting for the red fraction $\sigma_{f_{RS}}=0.14$, our $\sigma_\text{IG}$ increases from 0.10 to 0.23.
Furthermore, it is important to consider additional sources of scatter in observations, such as miscentering and projection effects \citep{Rozo2011}. Additionally, the choice of richness estimation methods employed by different cluster finders can also impact the scatter \citep{Rykoff2012,Adam2019}.

\section{conclusions}\label{sec06}
In this paper, we constrain the mass-richness (MR) relation of galaxy clusters with stellar mass-selected and magnitude-selected galaxies from two different hydrodynamic simulations, \gadgetx\ and \simba, from \MJ{THE300} project. We model the distribution of richness at a fixed cluster mass by a skewed Gaussian distribution (Equation \eqref{eq:Probability}) with a power-law scaling relation for the mean (Equation \eqref{eq:lnRichness}). Our main results are as follows:

\begin{itemize} 
	\item We display the fitting parameters and their variations with respect to the stellar mass threshold $M_{\star}$ and redshift $z$ in \autoref{fig:params}. The variation depends strongly on the baryon models when $M_{\star} \gtrsim 10^{10} \hMsun$. However, it is more stable with a lower $M_{\star}$. 
    For lower $M_\star$, the amplitude $A$ decreases with $M_\star$, increases with $z$, and these dependencies are independent. The slope $B$ and the scatter $\sigma_\text{I}$ remain constant with $M_\star$, while $B$ decreases with $z$ and $\sigma_\text{I}$ exhibits the opposite trend.
	
	\item We compare the fitting results from \gadgetx\ and \simba$\ $ in \autoref{fig:params}. For lower $M_\star$, \simba\ displays a stronger redshift evolution for $A$ and a negligible $\sigma_\text{I}$. The slope $B$ is consistent for the two simulations.
	
	\item We present the fitting parameters obtained from stellar mass-selected and magnitude-selected galaxies in \autoref{fig:Mstar_to_mag_GadgetX_abs_csst_i}. The relative difference in $\{A,B\}$ is within 5\%. However, $\sigma_\text{I}$ increases by 50\% in the magnitude-selected case. The relative difference between \gadgetx\ and \simba\ is basically propagated from $M_\star$ limit to magnitude limit.
	
	\item Additionally, we compare our skewed Gaussian distribution for the richness with the widely used log-normal distribution with a scatter given by Equation \eqref{eq:sigmaM}, as depicted in \autoref{fig:params4}, or a scatter $\sigma_\text{IG}$ given by Equation \eqref{eq:sigmaIG}, as depicted in \autoref{fig:SigmaIG}.
    The former exhibits a more intricate scatter with coupled and non-linear dependencies on halo mass, galaxy stellar mass threshold and reshift, while the latter demonstrates a mass dependence in $\sigma_{\rm IG}$, which has been overlooked in previous work.
	
	\item We provide the results of a 7-parameter fitting incorporating dependence on galaxy selection threshold and redshift for both the skewed Gaussian and log-normal distributions in \autoref{tab:Mstar_paramsZGadgetX} and \autoref{tab:Mstar_paramsZSIMBA}.
	
	\item Finally, to compare our findings with observational results in the literature, we combine our 7-parameter MR relation with the stellar mass-magnitude relation (Equation \eqref{eq:MstarMag}). The results are shown in \autoref{fig:ComparePapers_params}. After considering the red fraction, our results are consistent with the majority of literature at a pivot point, regarding richness, mass dependence, redshift dependence, and scatter.

\end{itemize}

Based on these results, we have established the MR relations from hydrodynamic simulations and demonstrated their applicability to actual observations. 
The differences between \gadgetx\ and \simba\ simulations offer valuable insights into the evolution of galaxies. 
While considering secondary halo properties, such as age and concentration, is expected to decrease the scatter in this relation  \citep{Hearin2013, Bradshaw2020, Farahi2020}, it is important to note that the intrinsic scatter $\sigma_\text{I}$ defined in this paper is more likely to originate from the large-scale environment with a strong dependence on baryon models. Further research and investigation (in preparation) are required to better understand the underlying physics.

One limitation of this study is the incompleteness of the low halo mass sample, especially at high redshifts, which can introduce biases to the result from the environmental effect. This is because the low-mass halos in \MJ{THE300} regions are mainly surrounded by the central clusters, which may cause some differences from those in other environments. However, we think the effect should be very small \citep{Wang2018}, especially for the galaxy number count with a large stellar mass cut.

Recently, there has been a lot of work using machine learning to estimate the mass of individual clusters\citep[e.g.][]{Ntampaka2015,Ntampaka2019,Yan2020,Ferragamo2023,deAndres2023}. This data-driven approach circumvents the need for dynamical or hydrostatic assumptions, effectively reducing the bias. Concurrently, numerous new mass proxies have emerged, including stellar mass \citep[e.g.][]{Andreon2012,Kravtsov2018,Pereira2020,Bradshaw2020}, stellar density profile \citep[e.g.][]{Huang2022} and intra-cluster light profile \citep[e.g.][and Contreras-Santos et al. in prep.]{Montes2019,Alonso2020}. Combining these studies, we expect that enhanced accuracy will be achieved in cluster cosmology and deeper comprehension will be brought to the formation and evolution of galaxies.

\begin{acknowledgments}

This work is supported by the National Key R\&D Program of China Grant No. 2022YFF0503404 and No. 2021YFC2203100, by the National Natural Science Foundation of China Grants No. 12173036 and 11773024, by the China Manned Space Project “Probing dark energy, modified gravity and cosmic structure formation by CSST multi-cosmological measurements” and Grant No. CMS-CSST-2021-B01, by the Fundamental Research Funds for Central Universities Grants No. WK3440000004 and WK3440000005, by Cyrus Chun Ying Tang Foundations, and by the 111 Project for``Observational and Theoretical Research on Dark Matter and Dark Energy" (B23042).
WC is also supported by the STFC AGP Grant ST/V000594/1, and the Atracci\'{o}n de Talento Contract no. 2020-T1/TIC-19882 granted by the Comunidad de Madrid in Spain. He also thanks the Ministerio de Ciencia e Innovación (Spain) for financial support under Project grant PID2021-122603NB-C21 and ERC: HORIZON-TMA-MSCA-SE for supporting the LACEGAL-III project with grant number 101086388. 

The simulations were performed at the MareNostrum Supercomputer of the BSC-CNS through The Red Española de Supercomputación grants (AECT-2022-3-0027, AECT-2023-1-0013), and at the DIaL -- DiRAC machines at the University of Leicester through the RAC15 grant: Seedcorn/ACTP317
\end{acknowledgments}

\appendix
\section{richness probability distributions}\label{sec:app1}
This section serves as a supplement to \autoref{sec31}.

\begin{figure*}
	\centering
	\includegraphics[width=1.\linewidth]{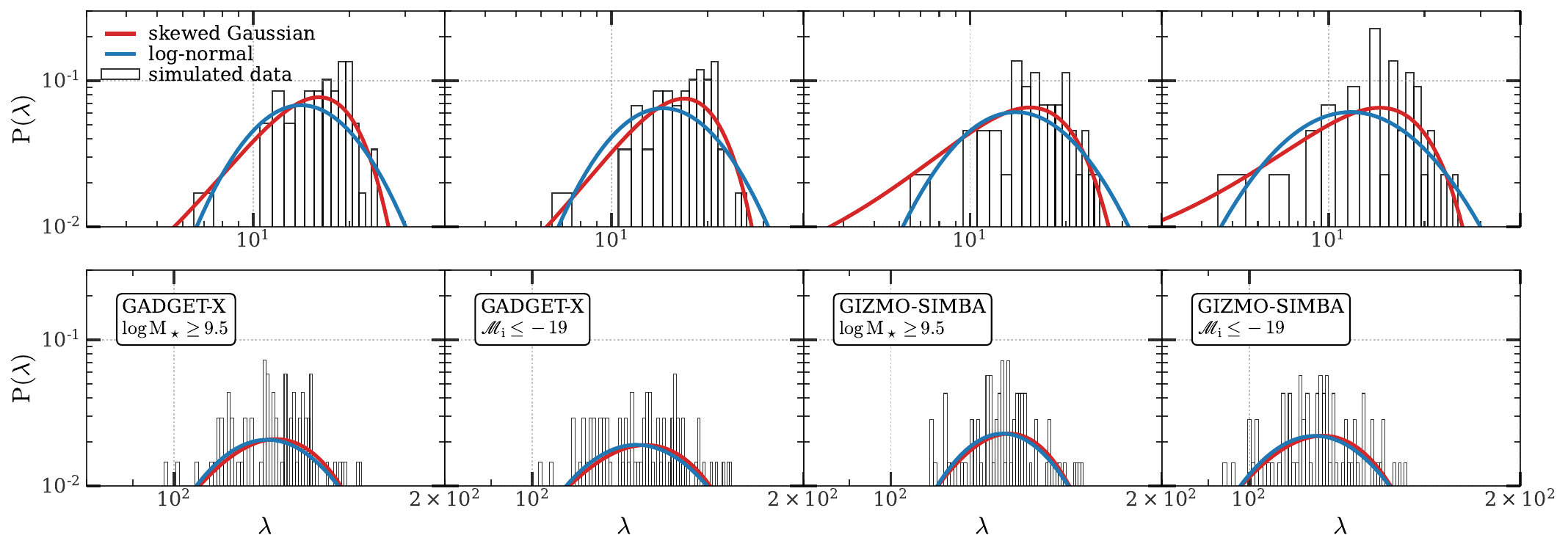}
	\caption{Similar to \autoref{fig:DataDistribution}. 
     Upper and lower panels correspond to mass bin $\log M[\hMsun]=[13.90,13.95]$ and $\log M[\hMsun]=[14.80,14.85]$, respectively.
     Each column represents different simulations, \gadgetx\ or \simba, as well as different galaxy selection thresholds, $\log M_\star [\hMsun] \geq 9.5 $ or $\mathscr{M}_i \leq -19$, as labeled in the legend.
     }
	\label{fig:DataDistribution4}
\end{figure*}

In \autoref{fig:DataDistribution4}, we show examples utilizing the skewed Gaussian function and the log-normal function to fit the individual richness probability distributions from both \gadgetx\ and \simba\ data in two mass bins at $z=0$. Both galaxy selections, i.e. the stellar mass $M_\star$ and magnitude $\mathscr{M}_i$, have been considered. The former function demonstrates better incorporation of low richness values, and both functions exhibit greater consistency in the larger mass bin.

\begin{figure}
	\centering
	\includegraphics[width=0.5\linewidth]{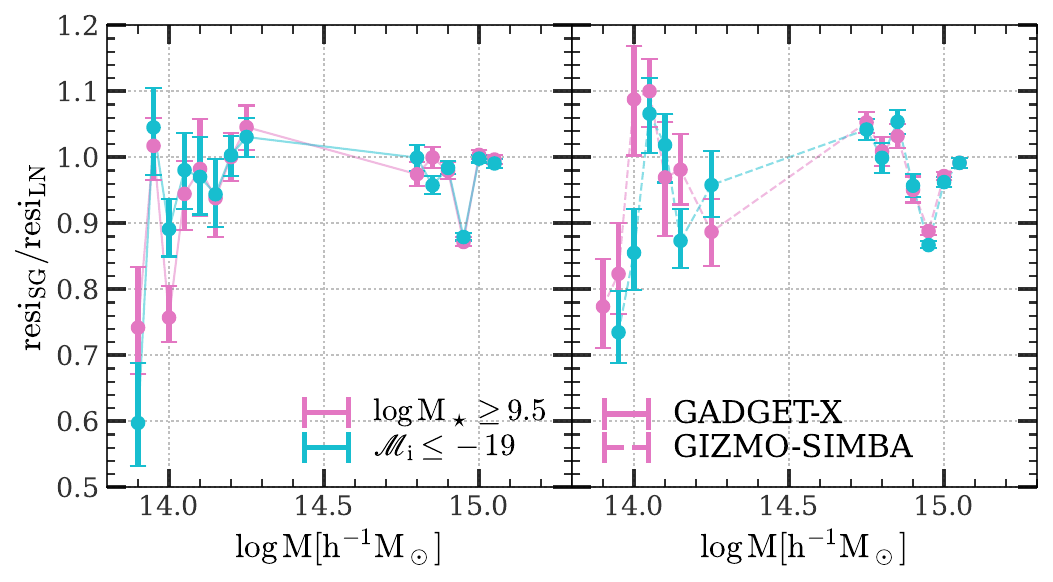}
	\caption{The ratio between residual using the skewed Gaussian function(resi$_\text{SG}$) and residual using the log-normal function(resi$_\text{LN}$).
    The left(solid lines) and right(dashed lines) panels correspond to \gadgetx\ and \simba\ respectively. 
    Different colors represent different galaxy selection thresholds, $\log M_\star [\hMsun] \geq 9.5 $(pink) or $\mathscr{M}_i \leq -19$(cyan).
     }
	\label{fig:DataDistributionResi}
\end{figure}

Additionally, we compute the ratio of residuals obtained from the skewed Gaussian function and the log-normal function. As illustrated in \autoref{fig:DataDistributionResi}, this ratio is largely less than 1, especially in the low mass bin.

\section{another form for the scatter} \label{sec:app2}
\begin{figure*}
	\centering
	\includegraphics[width=0.9\linewidth]{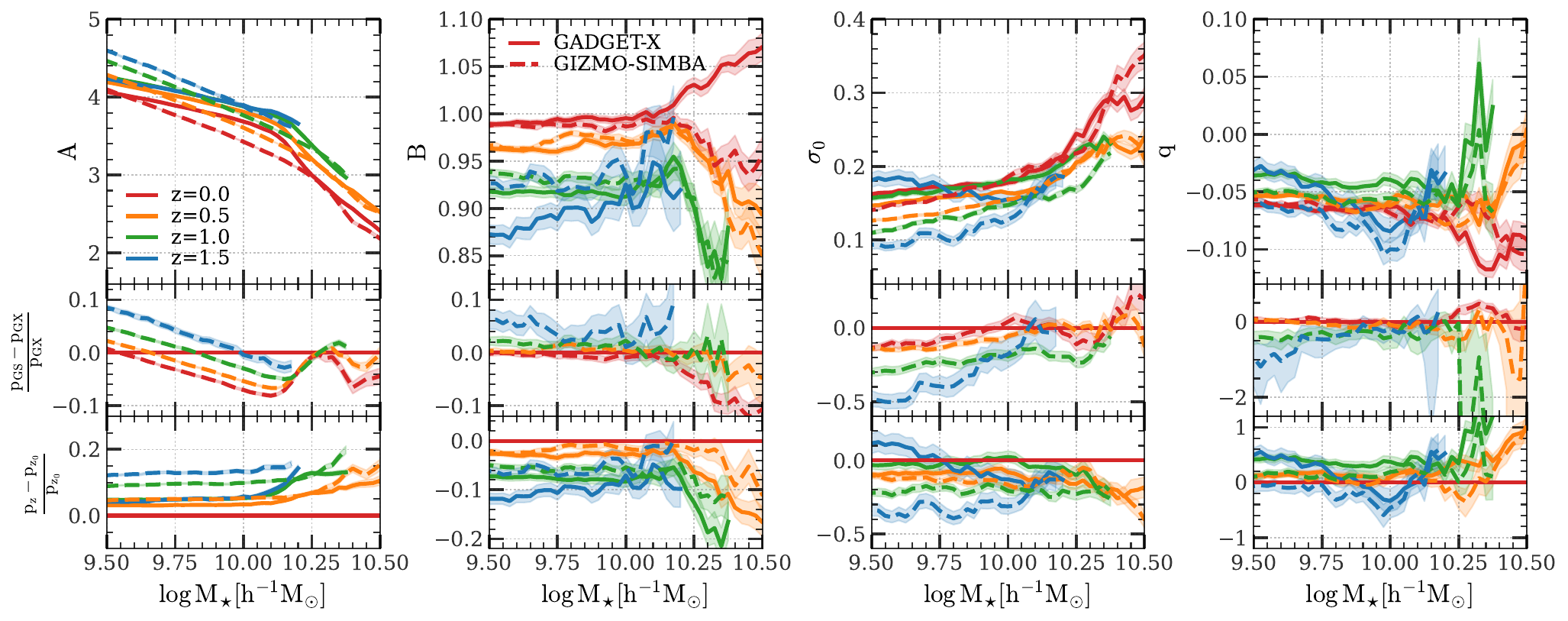}
	\caption{ Similar to \autoref{fig:params}, but for parameters $\{A,B,\sigma_0,q\}$.
	}
	\label{fig:params4}
\end{figure*}

There is another widely used assumption for the richness probability distribution \citep{Murata2018,Murata2019}, a log-normal form with a scatter that varies linearly with mass:
\begin{equation}\label{eq:sigmaM}
	\sigma_{\ln \lambda}=\sigma_0+q \ln \left(\frac{M}{M_{ {piv }}}\right), 
\end{equation}
involving more parameters than ours.
Nevertheless, we still repeat the procedure described in \autoref{sec41} to estimate the four parameters $\{A,B,\sigma_0,q\}$ for comparison. As depicted in \autoref{fig:params4}, the parameters $\{A,B\}$ are slightly different from our results, while the scatter $\sigma_0$, in particular, shows a strong dependence on $M_\star$.

\section{different bands}

\begin{figure}[htbp]
  \centering
  \begin{minipage}[b]{0.9\textwidth}
    \centering
    \includegraphics[width=\linewidth]{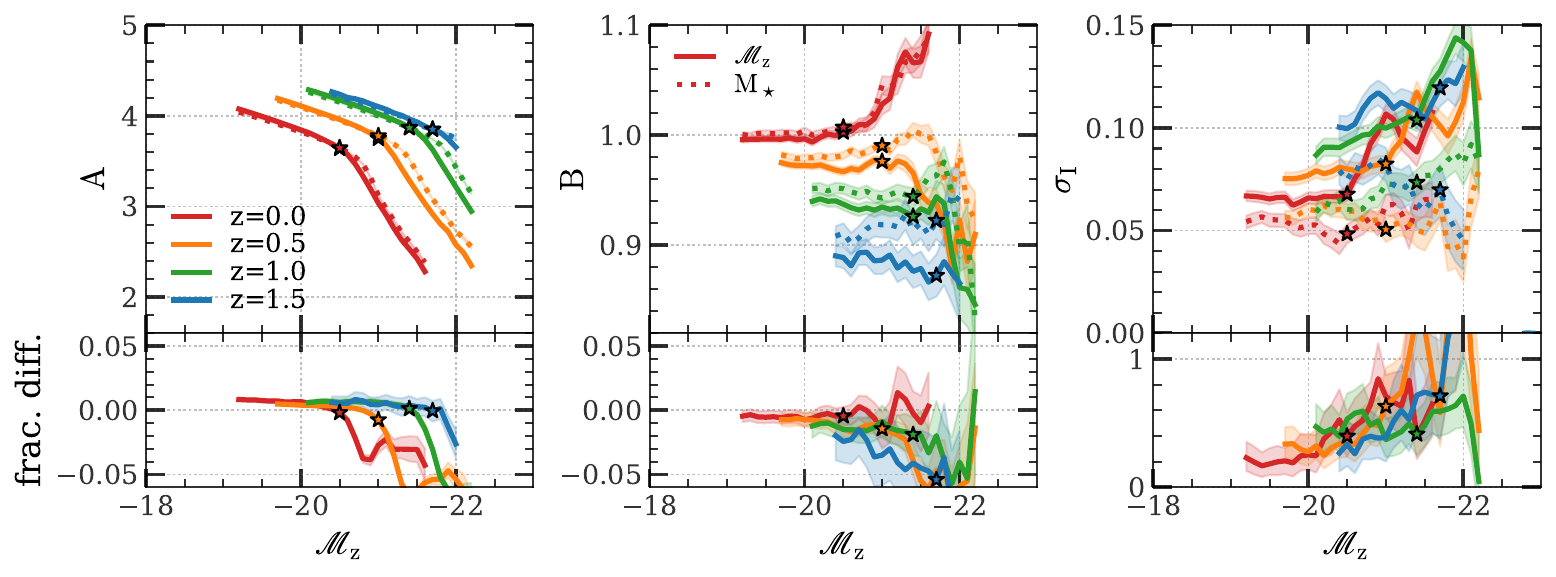}
  \end{minipage}%
  \begin{minipage}[b]{0.9\textwidth}
    \centering
    \includegraphics[width=\linewidth]{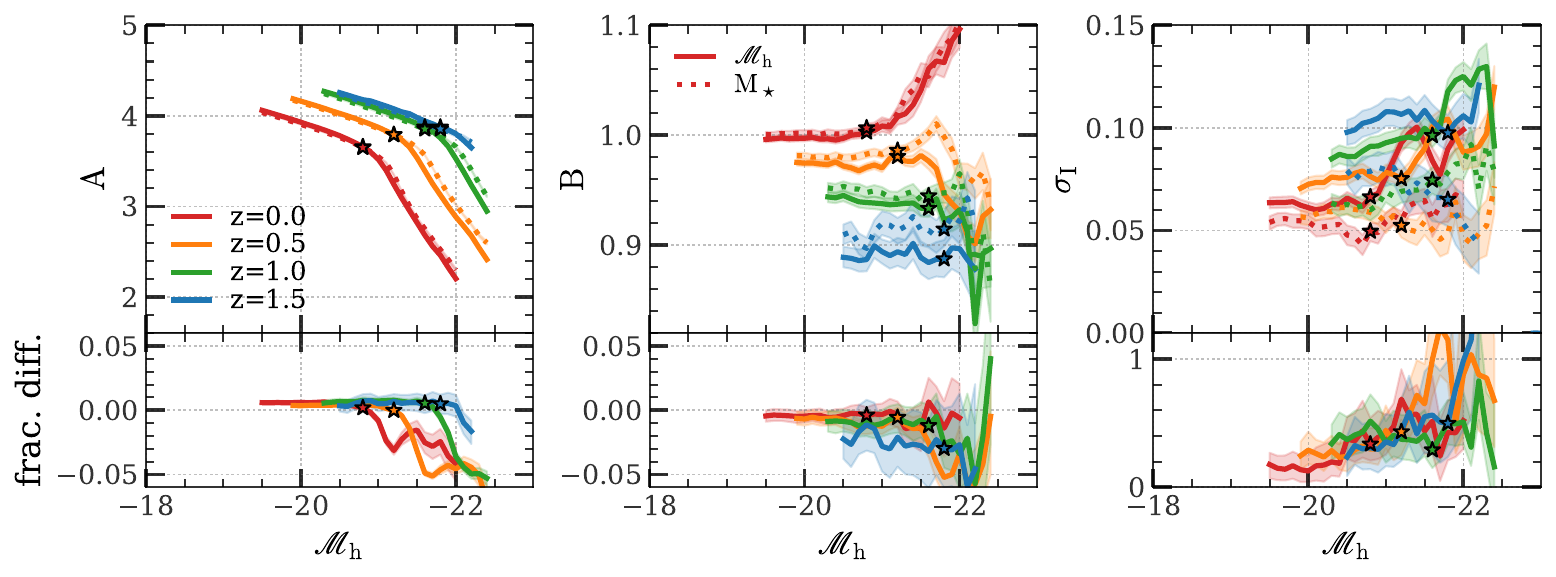}
  \end{minipage}
  \caption{Similar to \autoref{fig:Mstar_to_mag_GadgetX_abs_csst_i}. The upper panel is in CSST z-band. The lower panel is in Euclid h-band.}
  \label{fig:Mstar_to_mag_GadgetX}
\end{figure}

\begin{figure}[htbp]
  \centering
  \begin{minipage}[b]{0.9\textwidth}
    \centering
    \includegraphics[width=\linewidth]{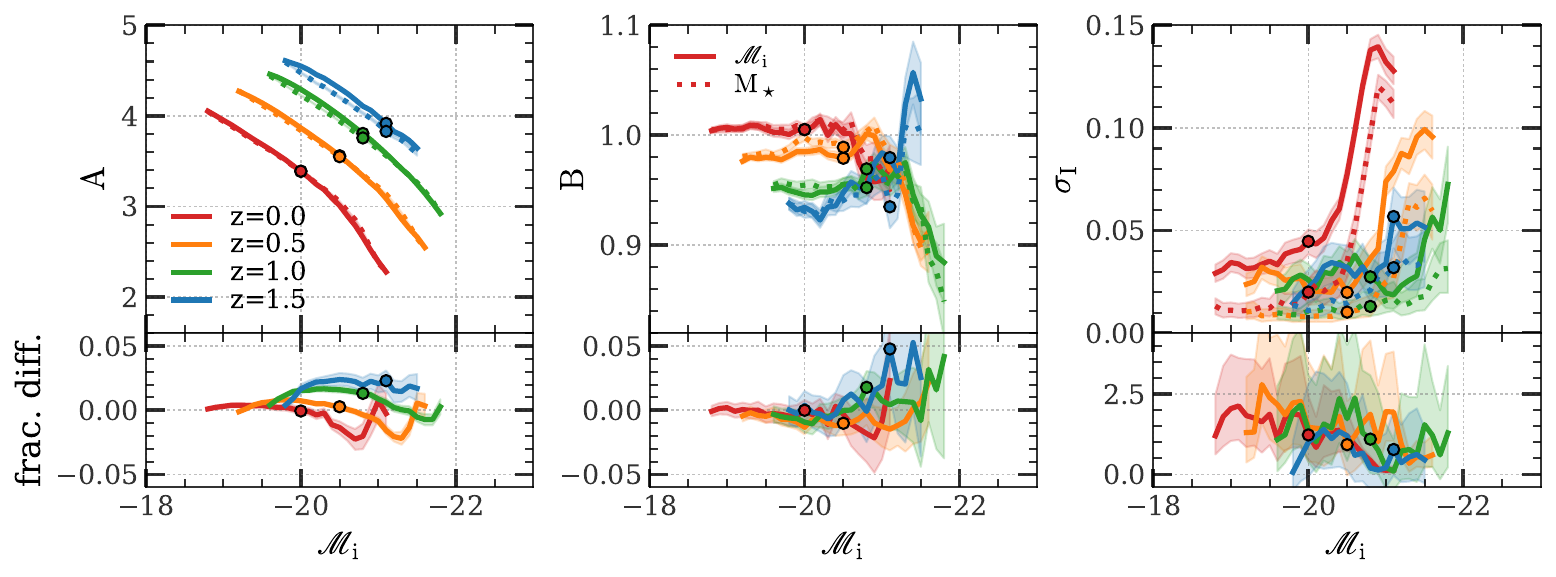}
  \end{minipage}%
  \begin{minipage}[b]{0.9\textwidth}
    \centering
    \includegraphics[width=\linewidth]{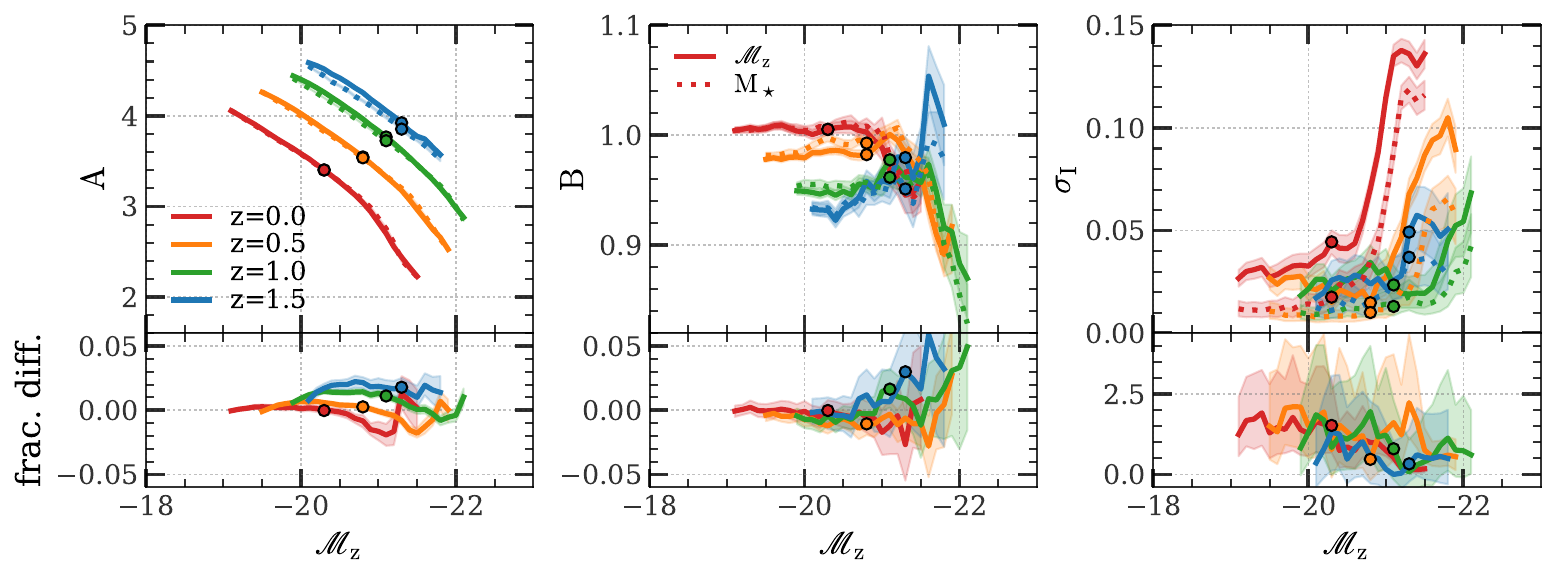}
  \end{minipage}
  \begin{minipage}[b]{0.9\textwidth}
    \centering
    \includegraphics[width=\linewidth]{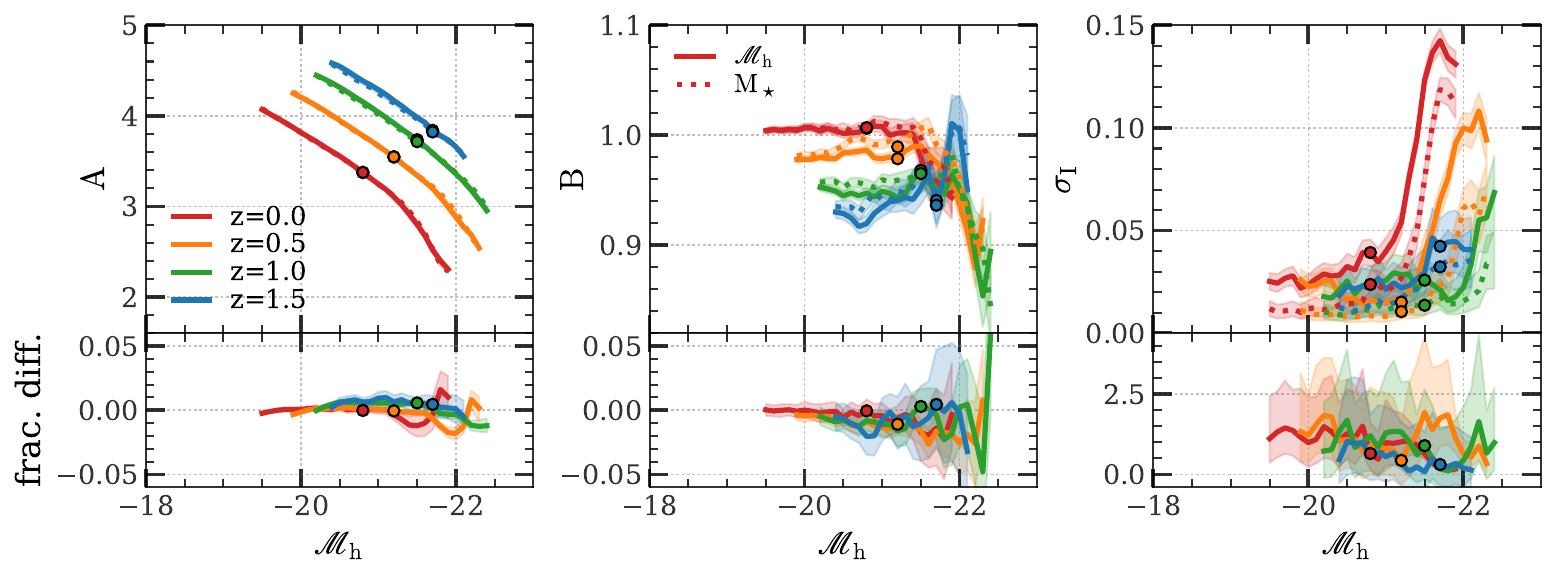}
  \end{minipage}
  \caption{Similar to \autoref{fig:Mstar_to_mag_GadgetX_abs_csst_i}, but for \simba. The upper panel is in CSST i-band. The middle panel is in CSST z-band. The lower panel is in Euclid h-band.}
  \label{fig:Mstar_to_mag_SIMBA}
\end{figure}

We fit the galaxy stellar mass-absolute magnitude $M_\star-\mathscr{M}$ relation in CSST-z band $\mathscr{M}_z$ and Euclid-h band $\mathscr{M}_h$ for both \gadgetx\ and \simba. The comparison results of the parameters are shown in \autoref{fig:Mstar_to_mag_GadgetX} and \autoref{fig:Mstar_to_mag_SIMBA}.

	\gadgetx:
	\begin{eqnarray}
		\ln M_\star=4.57 -0.90  \mathscr{M}_z -1.20  \times  \ln (1+z), \nonumber
		\\
		\ln M_\star=4.68 -0.88  \mathscr{M}_h -1.00  \times  \ln (1+z).
	\end{eqnarray}

	\simba:
	\begin{eqnarray}
		\ln M_\star=4.23 -0.92  \mathscr{M}_z -1.00  \times  \ln (1+z), \nonumber
		\\
		\ln M_\star=4.09 -0.91  \mathscr{M}_h -0.89  \times  \ln (1+z).
	\end{eqnarray}

\section{application in CSST}
We give an example employing the apparent magnitude threshold $m_i=25.9$ from \cite{Gong2019} to derive the MR relation. 
We use a simple relation, without K-correction or evolutionary correction,
\begin{eqnarray}
	\mathscr{M}_i = m_i-5\log \frac{D_L}{10 pc},
\end{eqnarray}
to convert the apparent magnitude $m_i$ to the absolute magnitude $\mathscr{M}_i$ at redshift $z=\{0, 1, 1.5\}$.
Equation \eqref{eq:MstarMag} is then utilized to obtain $M_\star$.
Next, by applying 7-parameters fitting results based on a log-normal distribution at each redshift bin $z=[0,0.5],\ [0.5,1],\ [1,1.5]$ from \autoref{tab:Mstar_paramsZGadgetX} and \autoref{tab:Mstar_paramsZSIMBA}, we obtain the MR relation 
 displayed as 3 parameters, as \autoref{tab:exampleCSST} shows. 
Note that the scatter here is the result of multiplying by 1.5, which is due to the transition from threshold $M_\star$ to $\mathscr{M}$ in \autoref{sec42}.
We would like to emphasize that this is merely a illustrative example and the threshold should be determined based on different cluster finders and richness estimators when practically applied, as demonstrated in \autoref{sec53}.

\begin{table}
	\caption{\label{tab:exampleCSST} Derived 3-parameters at the threshold $m_i=25.9$. $\{A,B,\sigma_\text{I} \}$ represents the skewed Gaussian distribution. $\{A,B,\sigma_\text{IG}\}$ represents the log-normal distribution.}
	\begin{ruledtabular}
		\begin{tabular}{l|cc|ccc|ccc} 
			$z$ &$\mathscr{M}_i$&$M_\star[ \hMsun]$&$A$&$B$&$\sigma_\text{I}$ &$A$&$B$&$\sigma_\text{IG}$\\
			\hline
			\multicolumn{9}{l}{\gadgetx}\\
			\hline
			0.5 & -16.39 & $1.82\times 10^{8}$ & 5.093 & 0.980 & 0.088 & 5.096 & 0.958 & 0.094 \\
            1.0 & -18.20 & $6.50\times 10^{8}$ & 4.766 & 0.952 & 0.094 & 4.765 & 0.916 & 0.114 \\
            1.5 & -19.27 & $1.29\times 10^{9}$ & 4.535 & 0.910 & 0.116 & 4.534 & 0.881 & 0.131 \\
			\hline
			\multicolumn{9}{l}{\simba}\\
			\hline
			0.5 & -16.39 & $2.38\times 10^{8}$ & 5.825 & 0.985 & - & 5.810 & 0.962 & - \\
            1.0 & -18.20 & $9.34\times 10^{8}$ & 5.192 & 0.958 & - & 5.190 & 0.928 & - \\
            1.5 & -19.27 & $1.98\times 10^{9}$ & 4.865 & 0.938 & - & 4.875 & 0.914 & - \\
		\end{tabular}
	\end{ruledtabular}
\end{table}

\bibliography{MRR}

\end{document}